\documentclass[twocolumn,showpacs,preprintnumbers]{revtex4}
\usepackage{amsmath}
\usepackage{graphicx}
\usepackage{dcolumn}
\usepackage{bm}

\begin{document}

\title{LDA+Gutzwiller Method for Correlated Electron Systems:
  Formalism and Its Applications}

\author{XiaoYu Deng, Lei Wang, Xi Dai, Zhong Fang}

\affiliation{Beijing National Laboratory for Condensed Matter
  Physics and Institute of Physics, Chinese Academy of Sciences,
  Beijing 100190, China}

\date{\today}

\begin{abstract}
  We introduce in detail our newly developed \textit{ab initio}
  LDA+Gutzwiller method, in which the Gutzwiller variational approach
  is naturally incorporated with the density functional theory (DFT)
  through the ``Gutzwiller density functional theory (GDFT)'' (which
  is a generalization of original Kohn-Sham formalism). This method
  can be used for ground state determination of electron systems
  ranging from weakly correlated metal to strongly correlated
  insulators with long-range ordering. We will show that its quality
  for ground state is as high as that by dynamic mean field theory
  (DMFT), and yet it is computationally much cheaper. In additions,
  the method is fully variational, the charge-density self-consistency
  can be naturally achieved, and the quantities, such as total energy,
  linear response, can be accurately obtained similar to LDA-type
  calculations. Applications on several typical systems are presented,
  and the characteristic aspects of this new method are clarified. The
  obtained results using LDA+Gutzwiller are in better agreement with
  existing experiments, suggesting significant improvements over LDA
  or LDA+U.
\end{abstract}

%\author{}
\pacs{71.15.-m, 71.27.+a, 71.15.Mb}% PACS, the Physics and Astronomy
\maketitle

\section{Introduction}
The density functional theory (DFT){\cite{DFT_HK,DFT_KS}} is very
successful in solid state physics and materials science.
First-principles calculations based on this theory, using the local
density approximation (LDA) or the generalized gradient approximation
(GGA), have been well developed and widely accepted as a powerful
theoretical tool for explaining and predicting ground state properties
and electronic structures of a large amount of materials such as
simple metals and band insulators.  However both the LDA and GGA fail
when they are applied to strongly correlated electron systems, a very
important class of materials in condensed-matter physics. These
materials contains unfilled d or f shells such as Cuprates,
Manganites, Ruthenates, Fe-pnictides,Plutonium as well as the Heavy
Fermion systems. In the last twenty years, many efforts have been made
to improve the situation, new methods, such as LDA+U{\cite{LDAU}},
self-interaction corrected (SIC) LDA{\cite{LDASIC}} and LDA plus
dynamical mean field (DMFT) theory{\cite{DMFT}}, have been proposed to
provide new computational tools for the quantitative study of the
strongly correlated materials.  Those methods are quite successful in
many aspects, nevertheless a method that is practically efficient and
can capture the key feature of the correlation effect as well is still
absent for the ground state studies.

One of the main features in correlated electron systems
is that although the electrons in those narrow 3d or 4f bands  are 
delocalized they still show some atomic features, which manifest itself
in the appearance of the Hubbard band and the enhancement of the
effective mass. In weakly correlated electron systems, electron states
are delocalized in real space, exhibiting nearly free electron
behavior which leads to good energy bands description. The
delocalization feature grants suitable electron density dependent
forms for correlation energy as presented in LDA and GGA since the
electron distribution is not far from homogeneous electron
gas. However, if electrons exhibit strong localization feature of the
atomic orbitals, it is better to describe the
electron states in real space.  The presence of strong on-site
correlations require proper treatment of atomic configurations, which
is orbital dependent and plays important roles in determining the
physical property in this case.  Methods such as LDA+U \cite{LDAU} and
LDA+DMFT \cite{DMFT} are proposed as remedies since this
orbital-dependent feature is absent in both LDA and GGA. These methods
start from similar Hamiltonians including on-site correlations but
operate in different ways. 

In LDA+U method, the on-site interaction
is treated in a static Hartree mean field manner, it is suited for
strongly correlated systems with long-range ordering, such as the AF
ordered insulators, but it fails for intermediately correlated
metallic systems. In DMFT method, the self energy which is purely local in
space is obtained in a self-consistent way, which make the LDA+DMFT 
method the most accurate
and reliable method now. However, the frequency dependent feature of
the self energy makes it very time consuming, and the
full charge density self-consistency, which is very important for the
accurate total energy calculation, is hard to be achieved.

Looking back to the progress of analytical treatment of strongly
correlated system, we can notice that the Gutzwiller variational
approach (GVA) has been proofed to be quite efficient and accurate
\cite{He3_vollhardt,Brinkman-Rice,fczhang_he3} for the ground state
studies of many important phenomena, i.e. the Mott transition,
ferromagnetism and superconductivity.  This approach was first
introduced by Gutzwiller to study the itinerant ferromagnetism in
systems with partially filled $d$ bands described by the Hubbard
Model\cite{Gut}. In this approach, a many body trial wave function was
proposed, in which the weights of unfavorable atomic configurations
are reduced according to the variational parameters. Both itinerant
and atomic features can be described spontaneously by this type of
wave functions. Thus, an unified description from weakly to strongly
correlated system can be built up by the GVA, this grants its
capability to accurately capture the essence of correlated systems.
Various techniques have been developed to formulate this approach
\cite{He3_vollhardt, vollhardt_MPLB,1d_exact_Gut,infinite_exact_Gut,
  multiband-G} for different model Hamiltonians. The reliability and
feasibility of GVA applied to correlated systems have been
demonstrated by these theoretical studies.

In this article, we will show that the GVA can be naturally combined
with the DFT. As the result, the LDA+Gutzwiller (simply called LDA+G
hereafter) method\cite{LDA+Gutzwiller} is proposed for practical
calculations of correlated electron systems. To understand the
formalism, we will show that a generalized Gutzwiller density
functional theory (GDFT) can be established following the same spirit
of Kohn-Sham (KS) formalism in the DFT. The GDFT itself is rigorous,
however, its exchange-correlation functional is unknown. By
introducing certain approximation to the exchange-correlation energy
in GDFT, the LDA+G method can be derived, very similar to the LDA or
LDA+U methods derived by approximation to the exchange-correlation
term in the KS formalism.  In order to show the validity and the
advantage of this method, we will demonstrate that GVA is as accurate
as DMFT for the ground state properties, but computationally much
cheaper. In addition, the present method is fully variational, which
guarantees that many of the important physical quantities, such as the
force or the linear response can be naturally obtained from the
variational principle. Detailed formalism of this method will be
explicitly introduced here, and we will also show that a fully charge
density self-consistent procedure can be carried out, which is quite
crucial for the total energy calculations.  Furthermore we will also
show that LDA+G method is easy to be implemented into the existing
codes, particularly if the LDA+U method is already available.

This paper is organized as follows. In Sec. II, GVA is introduced for
a multiband tight-binding Hamiltonian. Then we make detailed
comparison between GVA and DMFT results in Sec. III.  The combination
of GVA with DFT and its derivation from GDFT will be presented in
Sec. IV.  In Sec. V, we apply the method to several typical systems
and the results will be discussed. The proofs of some equations are
put in the Appendix.

\section{Gutzwiller Variational Approach}
We start with the GVA for the ground state of correlated electron
model systems. The detailed description of GVA has been presented by
many authors, here we refer to reference \cite{He3_vollhardt} for the
review.  For generality, we consider a model system with a set of
localized orbitals, such as $d$ or $f$ electrons, which can be
described quite generally by the multiband Hubbard model.  The
Hamiltonian reads \cite{multiband-G}

\begin{equation}
H=H_{0}+H_{int}=\sum_{i,j;\sigma ,\sigma ^{\prime }}t_{i,j}^{\sigma ,\sigma ^{\prime
}}C_{i\sigma }^{\dagger }C_{j\sigma ^{\prime
}}+\sum_{i}H_{i}  \label{eq:multibandH}
\end{equation}
and
\begin{equation}
H_{i}=\sum_{\sigma ,\sigma ^{\prime }(\sigma \neq \sigma ^{\prime })}{
\mathcal{U}}_{i}^{\sigma ,\sigma ^{\prime }}\hat{n}_{i\sigma }\hat{n}
_{i\sigma ^{\prime }} \label{eq:H_at}
\end{equation}
where $\sigma $ is combined spin-orbit index of localized orbitals
basis $ \{\phi _{\sigma }\}$ on site $i$, $\sigma =1,\ldots ,2N$ ($N$
is orbital number, e.g. $N=5$ for $d$ electrons). The first part is
just a tight-binding Hamiltonian extracted from LDA calculation, and
the second term is the local atomic on-site interaction in which only
density-density correlations are taken into account for
simplicity. For generalized on-site interactions, please refer to
Ref~\cite{multiband-G}.

We first examine the Hamiltonian in atomic limit (i.e, considering only
the $H_{i}$ term for single site). There are $2N$ different
spin-orbitals and each spin-orbital could be either empty or occupied,
thus totally $2^{2N}$ number of multi-orbital configurations
$|\Gamma\rangle$. $H_{i}$ is diagonal in the space casted by all
$|\Gamma \rangle $ configurations since the on-site interactions are
density-density type.

\begin{equation}
E_{i\Gamma}=\langle \Gamma |H_{i}|\Gamma \rangle =\sum_{\sigma ,\sigma
^{\prime }\in \Gamma }{\mathcal{U}}_{i}^{\sigma ,\sigma ^{\prime }}
\label{eq:UiI}
\end{equation}
$E_{i\Gamma}$ is the interaction energy of configuration
$|\Gamma\rangle$ for the $i$-th site.  (For general interactions, the
atomic part should be diagonalized, and the eigen vectors are linear
combination of $|\Gamma\rangle$). Of course those possible
configurations should not be equally weighted, and electrons tend to
occupy those configurations which has relatively lower energy. For
this purpose, we could construct projectors which project onto
specified configurations $|\Gamma\rangle$ on site $i$
\begin{equation}
\hat{m}_{i\Gamma}=\left\vert i,\Gamma \right\rangle \left\langle i,\Gamma
\right\vert  \label{eq:m_I_projection}
\end{equation}
with the normalization condition,
\begin{equation}
\sum_{\Gamma }\hat{m}_{i\Gamma }=1  \label{eq:local_completeness}
\end{equation}
since all the configurations \{ $|\Gamma \rangle $ \} form a locally
complete set of basis.

In Eq.~({\ref{eq:multibandH}}), if the interactions are not presented,
the ground state is exactly given by the Hartree uncorrelated wave
function (HWF) $|\Psi _{0}\rangle $, which is a single determinant of
 single particle wave functions. However,
after turning on the interaction terms, the HWF is no-longer an good
approximation, since there are many energetically unfavorable
configurations. In a physical view, to describe the ground state
better, the weights of those unfavorable configurations should be
suppressed. This is the main idea of Gutzwiller wave functions
(GWF). GWF $|\Psi _{G}\rangle $ is constructed by acting a
many-particle projection operator on the uncorrelated HWF.
\begin{equation}
\begin{split}
|\Psi _{G}\rangle &=\mathcal{\hat{P}}|\Psi _{0}\rangle \\
\mathcal{\hat{P}} &=\prod\limits_{i}\hat{P}_{i}=\prod\limits_{i}\sum_{
\Gamma }\lambda _{i\Gamma }\hat{m}_{i\Gamma }
\end{split}
\end{equation}

The role of projection operator $\mathcal{\hat{P}}$ is to adjust the
weight of each configuration through variational parameters
$\lambda_{i\Gamma }$ ($0\leq \lambda _{i\Gamma }\leq 1$). The GWF
falls back to non-interacting HWF if all $\lambda _{i\Gamma }=1$. On
the other hand, if $\lambda _{i\Gamma }=0$, the configuration $\Gamma$
on site $i$
will be totally removed. In this way, both the itinerant behavior of
uncorrelated wave functions and the localized behavior of atomic
configurations can be described consistently, and the GWF will give a
more reasonable physical picture of correlated systems than HWF does.

The evaluation of GWF is a difficult task due to its many-body nature.
There are many efforts in the literature, and the most famous one is
Gutzwiller approximation\cite{Gut}, introduced by Gutzwiller along
with his proposal of GWF. In this approximation, the inter-site
correlation effect has been neglected and the physics meaning was
discussed in Ref~\cite{He3_vollhardt} and Ref~\cite{Ogawa}.  The exact
evaluation of the single-band GWF in one dimension\cite{1d_exact_Gut}
and in the limit of infinite dimensions\cite{infinite_exact_Gut} were
carried out. It turns out that Gutzwiller approximation is exact in
the latter case. Extensions to multi-band correlated systems using
Gutzwiller approximation were carried out by J. B\"{u}nemann
\textit{et al.} \cite {multiband-G}. Meanwhile Gutzwiller
approximation was proofed to be equivalent to slave-boson
theories\cite{SB_Gut_kotliar, Coleman+Read, SB_Lechermann} on a
mean-field level for both one-band case\cite{1band_GW_SB} and
multi-band case\cite{Gebhard_2band,multi_GW_SB}.

The expectation value of Hamiltonian Eq.~(\ref{eq:multibandH}) in GWF is :
\begin{equation}
\langle H\rangle _{G}=\frac{\langle \Psi _{G}|H|\Psi _{G}\rangle }{\langle
\Psi _{G}|\Psi _{G}\rangle }=\frac{\langle \Psi _{0}|\mathcal{\hat{P}}H
\mathcal{\hat{P}}|\Psi _{0}\rangle }{\langle \Psi _{0}|\mathcal{\hat{P}}
^{2}|\Psi _{0}\rangle }  \label{eq:H_eff_gut}
\end{equation}
Using the Gutzwiller approximation, in the limit of infinite
dimensions, according to Ref~{\cite{multiband-G}} we have,
\begin{equation}
\begin{split}
\langle H\rangle _{G}=& \sum_{i\neq j;\sigma ,\sigma ^{\prime
}}t_{i,j}^{\sigma ,\sigma ^{\prime }}z_{i\sigma }z_{j\sigma ^{\prime
}}\langle C_{i\sigma }^{\dagger }C_{j\sigma ^{\prime }}\rangle
_{0}+\sum_{i;\sigma }\epsilon _{i\sigma }n_{i\sigma }^{0} \\
& +\sum_{i,\Gamma }E_{i\Gamma}m_{i\Gamma }
\end{split}
\label{eq:mgutzH}
\end{equation}
where $m_{i\Gamma}$ is the weight of configuration $\Gamma$,
\begin{equation}
m_{i\Gamma}=\langle \Psi _{G}|\hat{m}_{i\Gamma}|\Psi _{G}\rangle   \label{eq:m_I}
\end{equation}
and
\begin{equation}
\label{eq:zfac}
z_{i\sigma }=\sum_{\Gamma_i,\Gamma_i^{\prime}}\frac{\sqrt{m_{\Gamma _{i}}m_{\Gamma _{i}^{\prime }}}D_{\Gamma
    _{i}^{\prime}\Gamma _{i}}^{\sigma }}{\sqrt{n_{i\sigma }^{0}\left(
      1-n_{i\sigma }^{0}\right) }}
\end{equation}
with $D_{\Gamma ^{\prime }\Gamma }^{\sigma }=<\Gamma ^{\prime
}|C_{i\sigma }^{\dagger }|\Gamma >$, $0\leq z_{i\sigma }\leq 1$. (See
Appendix for details).

In order to understand the above Gutzwiller results properly, it is
better to compare it with the Hartree-Fock scheme. For this purpose,
here we give the Hartree-Fock expectation value of Hamiltonian
({\ref{eq:multibandH}}) using HWF $|\Psi_0\rangle$,
\begin{equation}
  \langle H\rangle _{0}=\sum_{i\neq j;\sigma ,\sigma ^{\prime
    }}t_{i,j}^{\sigma ,\sigma ^{\prime }}\langle C_{i\sigma }^{\dagger
  }C_{j\sigma ^{\prime }}\rangle
  _{0}+\sum_{i;\sigma }(\epsilon _{i\sigma }+\Delta\epsilon_{i\sigma})n_{i\sigma }^{0}+C
\label{E_HFW}
\end{equation}
where C is a constant , and $\Delta\epsilon_{i\sigma}$, which is
proportional to interaction strength $U$, is a correction to the
on-site energy (level shift) introduced by the static mean field
treatment of the interaction term.

Comparing Eq. (\ref{eq:mgutzH}) with (\ref{E_HFW}), now it is clear
that the main differences between the Gutzwiller and the Hartree
approaches are: (1) There are orbital-related factors $z_{i\sigma }$
in the former associated with the hopping terms, which describe the
renormalization of kinetic energy, while the kinetic energy in the
Hartree approach is not renormalized; (2) The interaction energy in
the Gutzwiller approach is not simply scaled with the interaction
strength $U$, but it is related to the configuration weights. While in
the Hartree approach, the presence of interaction term will contribute
simply to the on-site energy correction in proportional to $U$ after
the mean field treatment. 

The total energy under the GWF can be obtained by minimizing
Eq.~({\ref{eq:mgutzH}}) with respect to configuration weights
$m_{i\Gamma}$, which now in fact are variational parameters. Since
more variational parameters are presented in this approach, the
obtained ground state total energy is much better than that in HWF. In
other words, by using the GWF, the obtained ground state total energy
is further reduced due to the reduction of interaction energy, but in
the cost of kinetic energy. The balance of two ({\it gain} and {\it
  cost}) is achieved by the energy minimization with respect to
variational parameters.

For the convenience of our following discussions, here we would like
to generalize the formalism and make several definitions.  Any
operator $\hat{A}$ acting on the GWF, can be mapped to an
corresponding Gutzwiller effective operator $\hat{A}^G$ which acts on
the HWF (rather than GWF), requiring that its expectation values is
kept as the same,
\begin{equation}
  \langle \Psi _{G}|\hat{A}|\Psi _{G}\rangle =
\langle \Psi _{0}|\mathcal{\hat{P}}^\dag \hat{A} \mathcal{\hat{P}} |\Psi _{0}\rangle 
=\langle \Psi _{0}|\hat{A}^{G}|\Psi _{0}\rangle 
\end{equation}
here we have,
\begin{equation}
A^G=\mathcal{\hat{P}}^\dag A \mathcal{\hat{P}}
\label{definition_G}
\end{equation}

If the operator $\hat{A}$ is a single-particle operator, such as
$\hat{A}_0=\sum_{ij,\sigma\sigma^\prime}A_{ij}^{\sigma\sigma^\prime}C_{i\sigma}^\dag
C_{j\sigma^\prime}$ (where $A_{ij}^{\sigma\sigma^\prime}=\langle
i\sigma|\hat{A_0}|j\sigma^\prime\rangle$), then similar to the above
procedure for the evaluation of kinetic energy in
Eq.~({\ref{eq:mgutzH}}), its Gutzwiller effective operator (in
Gutzwiller approximation) can be written as,
\begin{equation}
\begin{split}
 \hat{A}_0^{G} 
   = \sum_{i j;\sigma \sigma ^{\prime }}A_{ij}^{\sigma\sigma'} z_{i\sigma}
z_{j\sigma'}  C_{i\sigma }^{+}C_{j\sigma ^{\prime }}
   +\sum_{i,\sigma }(1-z_{i\sigma }^{2})  C_{i\sigma}^{+}C_{i\sigma }
 A_{ii}^{\sigma\sigma}
\end{split}
\label{eq:oneparticleoperator}
\end{equation}
where again $z_{i\sigma}$ are orbital dependent renormalization
factors, which are determined through the configuration weights
presented in the projector $\mathcal{\hat{P}}$. Here please note that
the diagonal term and the hopping term should be treated separately
(see the Appendix).

Following the above general definition, we can now define an
Gutzwiller effective Hamiltonian $H^G$ which acts on HWF,
\begin{equation}
H^G=H_0^{G}+H_{int}^G
\end{equation}
such that the following equation holds,
\begin{equation}
  \langle \Psi _{0}\vert H^G\vert\Psi _{0}\rangle= \langle \Psi _{G}\vert H\vert\Psi _{G}\rangle =E_G
\end{equation}
here the kinetic part $H_0^G$ can be written out according to
Eq.~(\ref{eq:oneparticleoperator}), and the interaction part is
\begin{equation}
  \begin{split}
H_{int}^G=\sum_{i;\Gamma}E_{i\Gamma}\hat{m}_{i\Gamma}  \label{eq:Heff}    
  \end{split}
\end{equation}
for the density-density type interaction as discussed above.

Now we are coming to a stage that we can solve the Gutzwiller problem
easily through energy minimization. In practice, the minimization
procedure will be done iteratively with each loop being divided into
two steps. The first step is to fix the Gutzwiller variational
parameters $m_{\Gamma}$ and find the optimal HWF.  As we know the
$H^{G}$ for given $m_{\Gamma}$, which is non-interacting, this step
can be easily done by diagonalize it and fill the corresponding bands
up to the Fermi level.  Then in the next step, we will fix the HWF and
optimize the energy respect to all the Gutzwiller variational
parameters $m_{\Gamma}$.  The explicit equation can be written as:
\begin{equation}
  \begin{split}
&    \frac{\partial E_G}{\partial m_{i,\Gamma}}=\sum_{j,j\neq i} [\sum_{\sigma\sigma'}t_{i,j}^{\sigma ,\sigma ^{\prime }}\frac{\partial z_{i,\sigma }}{\partial m_{i,\Gamma}}z_{j,\sigma ^{\prime
}}\langle C_{i\sigma }^{\dagger }C_{j\sigma ^{\prime }}\rangle_{0} \\
&+\sum_{\sigma\sigma'}t_{j,i}^{\sigma ,\sigma ^{\prime }}z_{j,\sigma ^{\prime}}\frac{\partial z_{i,\sigma' }}{\partial m_{i,\Gamma}}
\langle C_{j\sigma }^{\dagger }C_{i\sigma ^{\prime }}\rangle_{0} ]
+E_{i\Gamma}=0
  \end{split}
\end{equation}
In this second step of calculations, for the lattice model with
crystal periodicity, usually additional constrains can be adopted: (1)
There is no site-dependency for $z_{i\sigma}$ factors and occupation
number $n_{i\sigma}$, i.e, $z_{i\sigma}=z_{j\sigma}$ and
$n_{i\sigma}=n_{j\sigma}$; (2) The charge on each orbital should be
kept to be the same as that obtained by HWF (for pure density-density
interaction as discussed in \cite{multiband-G}), in other words we
have:
\begin{equation}
  \sum_{\Gamma }\left\langle \Gamma \right\vert C_{i\sigma }^{\dag }C_{i\sigma
  }\left\vert \Gamma \right\rangle m_{i,\Gamma }=n_{i\sigma}
  =n_{i\sigma}^0=\langle\hat{n}\rangle_0
\end{equation}
When all $m_{i\Gamma}$ are obtained, go back the first step to
construct an new effective Gutzwiller Hamiltonian again. By this
recursive method, all parameters $m_{i\Gamma}$ and $\vert \Psi_0 \rangle$
can be obtained self-consistently.

Typically, the second step of variations, i.e. the optimization of
$m_{i\Gamma}$, is not so easy for multi-band systems, because a large
number of non-linear equations need to be solved
spontaneously. Fortunately, following the steps described in our
previous publication~\cite{Xi-Dai}, we are able to transfer the
non-linear equations into linear equations set, and furthermore a so
called "adiabatic solution searching" procedure can be adopted. Those
techniques will greatly reduce the computational cost and stabilize
the calculations.

\section{Comparison of Gutzwiller Approximation with Dynamical Mean
  Field Theory}

In this section, we will compare the results obtained by the GVA and
that by the dynamical mean field theory (DMFT) for the single and two
bands Hubbard model. With the careful comparison of kinetic energy,
interaction energy and quasi-particle spectrum, we are going to
clarify the following important issue: Can the Gutzwiller
approximation (GA) capture the important \textquotedblleft incoherent
motion\textquotedblright\ of the correlated-electrons or not?  This
problem is considered to be the biggest shortcoming of GA, which
prevents it to be widely used in the first-principles calculations of
strongly correlated materials. As we will show below, the GA can
definitely capture the effect of {}\textquotedblleft
incoherent\textquotedblright\ motion in the ground state by its
multi-configuration nature, which leads to very good agreement with
the DMFT ground state results for both the kinetic and interaction
energies.  While the situation is not as good for the excited states,
since the variational parameters in the GA are determined by
optimizing only the ground state energy, not that of excited
states. Therefore, GA is a much better approximation for the ground
state than for excited states. Within the frame of GA, it is difficult
to construct the high energy excited states corresponding to the upper
and lower Hubbard bands. That is why in the green's function obtained
by GA, we only have quasi-particle part and no Hubbard bands. While
the problem only exists for the high energy excited states not for the
ground state and the low energy quasi-particle states.

We start from the multi-band Hubbard model (\ref{eq:multibandH}), for
the clarity we only keep the intra-orbital hoping $t_{i,j}^{\sigma
  ,\sigma ^{\prime }}=t_{i,j}\delta _{\sigma ,\sigma ^{\prime }}$ and
neglect the on-site energy $\epsilon _{i;\sigma }$ in the following
(restoring them does not change the conclusions). To describe
quasi-particle states, an important physical quantity is
$Z$-factor. Actually, there are two different definitions of $Z$ in
literature . The first one is the renormalization factor of the
effective band width for the quasi-particles, the second one is the
weight of the coherent part in the electron green's function near
Fermi surface. As we will show below, in GA the $Z$-factors obtained
by the above two definitions match each other, while in DMFT they are
quite different. In the following comparison, we compute $Z$ within
DMFT by $Z=(1-%
\frac{\partial \Re \Sigma ^{R}}{\partial \omega }|_{\omega =0})^{-1}$,
which is quasi-particle weight. While in GA calculation we define
$Z=z^{2}$.

Under Gutzwiller approximation, the quasi-particle states and quasi-hole
states can be expressed as\cite{gutzwiller_qp}

\begin{equation*}
|\Phi _{k\sigma }^{p/h}\rangle =\left\{
\begin{array}{cc}
\mathcal{\hat{P}}C_{k\sigma }^{\dagger }|\Psi _{0}\rangle  & \text{for }%
\varepsilon _{k\sigma }>\mu _{F} \\
\mathcal{\hat{P}}C_{k\sigma }|\Psi _{0}\rangle  & \text{for }\varepsilon
_{k\sigma }<\mu _{F}%
\end{array}%
\right.
\end{equation*}%
With the above trial wave function, the excitation energy can be
calculated as

\begin{equation}
\pm E_{k\sigma }^{p/h}=\frac{\langle \Phi _{k\sigma }^{p/h}|H|\Phi _{k\sigma
}^{p/h}\rangle }{\langle \Phi _{k\sigma }^{p/h}|\Phi _{k\sigma
}^{p/h}\rangle }-E_{G}  \label{excited_en}
\end{equation}
for quasi-particle (quasi-hole) excitations. The above equation can be
evaluated by GA as shown in Appendix, which leads to a simple
expression of Green's function,
\begin{equation*}
G_{k\sigma }^{coh}(i\omega )=\frac{\gamma _{k\sigma }^{2}}{i\omega
-z_{k\sigma }^{2}\left( \varepsilon _{k\sigma }-\mu _{F}\right) }
\end{equation*}
with
\begin{equation*}
\gamma _{k\sigma }^{2}=\left\{
\begin{array}{cc}
\left\vert \langle \Phi _{k\sigma }^{p}|C_{k\sigma }^{\dagger }|\Psi
_{G}\rangle \right\vert ^{2} & \text{for }\varepsilon _{k\sigma }>\mu _{F}
\\
\left\vert \langle \Phi _{k\sigma }^{h}|C_{k\sigma }|\Psi _{G}\rangle
\right\vert ^{2} & \text{for }\varepsilon _{k\sigma }<\mu _{F}%
\end{array}%
\right.
\end{equation*}%
being the weight of the coherent part spectrum, which can also be
evaluated to be equal to $z_{\sigma }^{2}$ under GA as shown in
Appendix. Therefore within GA the quasi-particle weight coincident
with the renormalization factor of the kinetic energy, thus dynamical
informations are captured by variational approach.

\begin{figure}[tbp]
\includegraphics[clip,width=8cm] {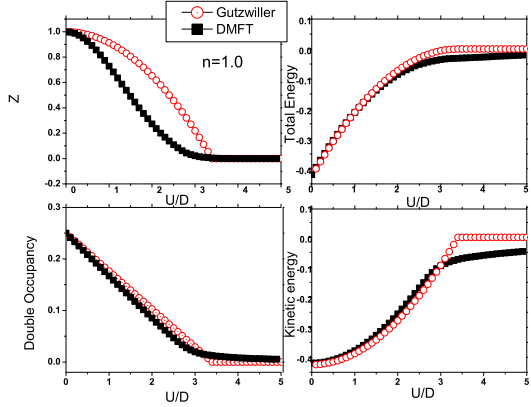}
\caption{Comparisons of calculated Z-factor, total energy, double
  occupancy and kinetic energy of single band Hubbard model with
  half-filling. In GA the band renormalization factor and
  quasi-particle weight coincident. Z-factor from DMFT denotes
  quasi-particle weight. Double occupancy is $<n_{\uparrow
  }n_{\downarrow }>$}.
\label{1.0}
\end{figure}

\begin{figure}[tbp]
\includegraphics[clip,width=8cm] {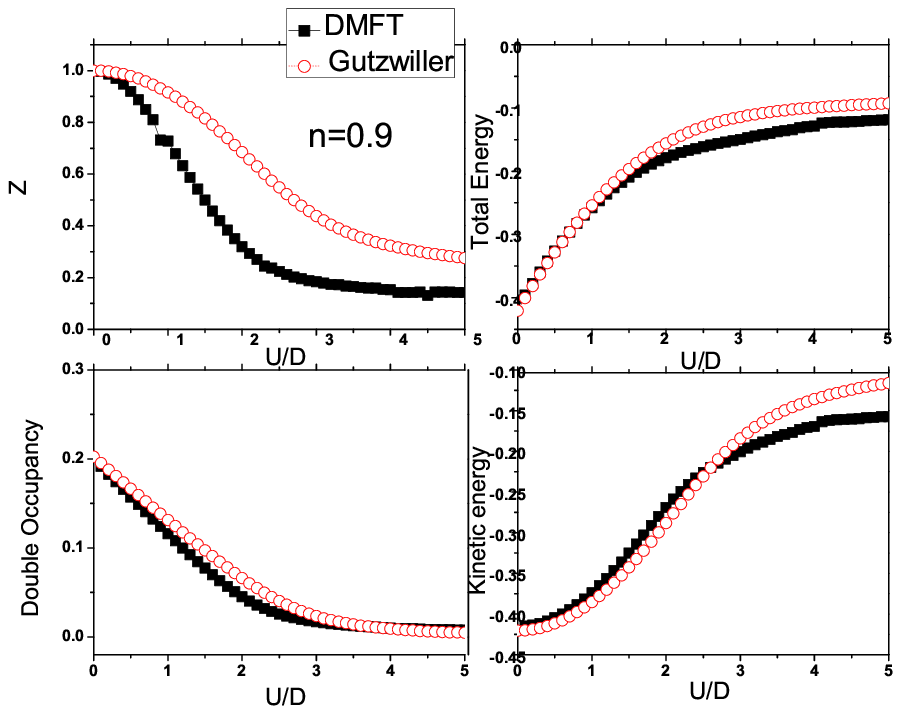}
\caption{Comparisons of calculated Z-factor, total energy, double
  occupancy and kinetic energy of single band Hubbard model with
  occupation number $n$=0.9.}
\label{0.9}
\end{figure}

\begin{figure}[tbp]
\includegraphics[clip,width=8cm] {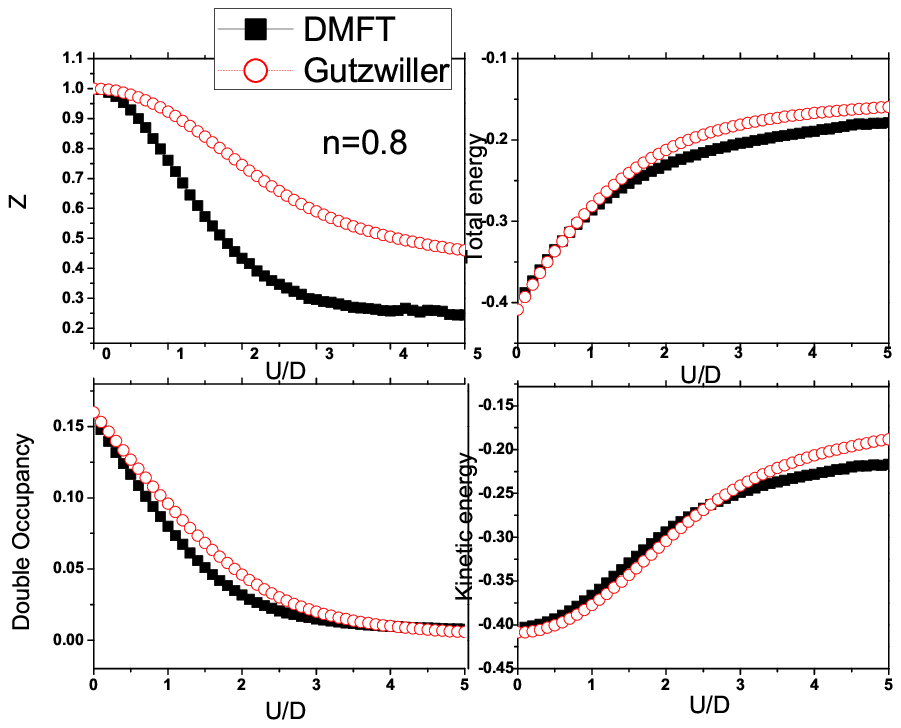}
\caption{Comparisons of calculated Z-factor, total energy, double
  occupancy and kinetic energy of single band Hubbard model with
  occupation number $n$=0.8.}
\label{0.8}
\end{figure}

\begin{figure}[tbp]
\includegraphics[clip,width=8cm] {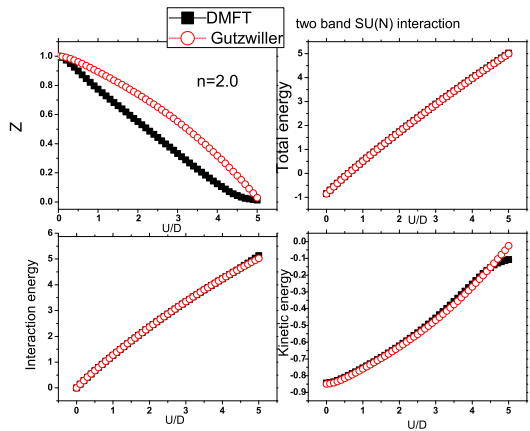}
\caption{Comparisons of calculated Z-factor, total energy, interaction
  energy and kinetic energy of two band Hubbard model with SU(N)
  interaction.}
\label{2band}
\end{figure}

In Fig. \ref{1.0}-\ref{2band}, we compare the kinetic energy,
interaction energy and $Z$ -factor from GA and DMFT calculations. We
choose the non-interaction density of state to be $\rho (\varepsilon
)=\frac{2}{\pi D}\sqrt{D^{2}-\varepsilon ^{2}}$, which corresponds to
Bethe lattice with infinite connectivity.  Anderson impurity model in
DMFT is solved by Lanczos method, which gives essentially exact
results. As could be seen in Fig. \ref{1.0}$\sim $\ref{2band}, GVA
captures ground state energies quite well for almost all correlation
strength and band fillings. Although as a variational approach it
targets at total energy and does not ensure the correctness of kinetic
and interaction energy in principle, we still observe quite good
coincidence of kinetic and interaction energy respectively. We also
notice that as the band degeneracy increase there is further
coincident between GA and DMFT results as shown in the Fig.4 for two
band case.

For half-filling $n=1.0$ where there is Mott insulator transition for
large $U$, the introduction of GA further neglects spatial correlation
and under-estimate the absolute value of kinetic energy
(Fig. \ref{1.0}). In Mott phase, GA gives vanishing double occupancy,
while DMFT result always shows finite double occupancy due to spatial
fluctuation. The flaw at large $U$ could be traced back to the fact
that the starting wave function of Gutzwiller projection is
uncorrelated Fermi liquid state $|\Psi_{0}\rangle$. This fact is also
part of the reasons for the important shortcoming of GA: it
miscaptures the high energy excited states (seen from the overestimate
of $Z$-factor in Fig. \ref{1.0}$\sim $\ref{2band} for large $U$). For
all interaction strengths and band fillings, $Z$-factor from GA is
larger than DMFT treatment, \textit{i.e.,} the method gives more
weight to low energy coherent part. Nevertheless, we recently
demonstrated~\cite{TMA} that this overestimation of $Z$-factor can be
further corrected by properly taking into account the contribution from
excited states (namely the incoherent motion of electrons).

To conclude this section, we make the assertion that as a cheap tool
for correlated systems GA has fairly good energy resolution,
particularly good for the ground state total energy with dynamical
information included, but careful must be taken when dynamical
information associated with high energy excitations is trying to be
extracted from the GA results.

\section{Combining DFT with Gutzwiller Variational Approach}

In this section, we will discuss how we can combine the DFT with the
GVA. The discussions will be separated into three parts. The detailed
formalism of LDA+G method are explicitly derived in the first part,
which are used in realistic calculations.  In the second part, we
first introduce a general Gutzwiller density functional theory (GDFT),
and then we derive the LDA+G formalism from the firm base of GDFT. In
such a way, we demonstrate the rigidity of this method. Finally, in
the third part, we will discuss the on-site interactions and the
double counting term.

\subsection{Formalism of LDA+Gutzwiller Method}

As we discussed above, the strong on-site correlation is
underestimated in LDA. For those strongly correlated materials, in
which the correlations play very important roles in determine the
electronic structure, this underestimation may lead to qualitative
mistakes. One common procedure to overcome this problem is that we
treat the interactions more explicitly on top of LDA level, just like
what has been done in LDA+U or LDA+DMFT schemes. The starting
effective Hamiltonian is usually written as:
\begin{eqnarray}
  H & = & H_{LDA}+H_{int}-H_{dc}
\label{LDA++}
\end{eqnarray}
where $H_{LDA}$ is the LDA part Hamiltonian extracted from the
standard LDA calculation, $H_{int}$ is the on-site interaction term,
and $H_{dc}$ is the double counting term representing the average
orbital independent interaction energy already included by LDA.

Both LDA+U and LDA+DMFT methods start from the same Hamiltonian as
shown above, however they treat the problem in different ways. In the
LDA+U scheme, Hartree like mean field approximation is used to solve
the above Hamiltonian, which can capture the orbital dependent physics
(which is absent in LDA), but the dynamical correlation is still not
included.  While in the LDA+DMFT method, the purely local self energy
is evaluated by solving an effective quantum impurity model mapped
from the original lattice model. With the frequency dependent self
energy, not only the ground state properties but also the dynamical
response around the equilibrium can be considered by LDA+DMFT. Because
of the frequency dependency of the self energy, the LDA+DMFT is quite
expensive in computational time. In many applications, we are only
interested in the ground state properties and it is quite important to
develop a new computational method for correlation materials, which is
as fast as LDA+U and can capture the dynamical correlation effect as
well for the ground state.

As we have proposed in the previous paper \cite{LDA+Gutzwiller}, an
alternative way to solve the problem is to use Gutzwiller wave
function rather than single determinant Hartree wave function. This
approach is much cheaper than DMFT, but its quality is as good as DMFT
for the ground state determination (as been shown in the last
section), because it can capture the dynamical correlation effect due
to the multi-configuration nature of the Gutzwiller wave
function. More importantly, this approach is fully variational, and
can be easily combined with the DFT as will be discussed below.

Now the goal is to solve the Hamiltonian~(\ref{LDA++}) by the GVA. For
this purpose, we need to discuss the Hamiltonian in more detail.
Since the problem to be addressed here is generally orbital-dependent,
the effective Hamiltonian should be written in a set of complete
orbital basis, which are always available, such as wannier functions
or atomic orbitals. These orbitals can be denoted by $\vert
i\alpha\rangle$, in which $i$ is site index, $\alpha$ is spin-orbital index
and $C_{i\alpha}^{\dagger}$ is the corresponding creation operator.

Following the basic idea of LDA+U or LDA+DMFT approaches, the
$H_{LDA}$ term in the effective Hamiltonian~(\ref{LDA++}) is regarded
as single-particle operator, it can be therefore expressed in terms of
a complete set of orbitals as
\begin{equation}
\begin{split}
  H_{LDA}=&\sum_{ij,\alpha\alpha'}t_{ij\alpha\alpha'}C_{i\alpha}^{\dagger}C_{j\alpha'} \\
t_{ij\alpha\alpha'}=&\langle i\alpha \vert H_{LDA} \vert j\alpha^\prime \rangle
\end{split}
\end{equation}
Suppose all the orbitals on the same site are correlated, and the
interaction term can be written as
 \begin{equation}
   H_{int}=\sum_{i\alpha\alpha^{\prime}(\alpha\neq\alpha^{\prime})}
   {\mathcal{U}}_{i}^{\alpha,\alpha^{\prime}}
   \hat{n}_{i;\alpha}\hat{n}_{i;\alpha^{\prime}}
\end{equation}
in which $\hat{n}_{i\alpha}=C_{i\alpha}^{\dagger}C_{i\alpha}$. Now it
is easy to see that the effective Hamiltonian~(\ref{LDA++}) has the
same form as that shown in Hamiltonian~(\ref{eq:multibandH}). (Please
note the double counting term only contribute to a constant uniform
shift, and has no orbital-dependency, as will be addressed in the
later part). Then following the steps discussed in the section I, we
will be able to solve the problem. This is the scheme used in most of
post-LDA techniques, where a tight-binding fit to LDA results are
first obtained, and then local orbital dependent interaction terms 
are implemented and the
problem with interaction should be solved by some many-body techniques. We
can therefore call the above procedures as post-LDA plus Gutzwiller
approaches, which has been recently used for several
examples~\cite{GW_Pu,GW_Ni,GW_NaCoO}.

However, our intention is to develop a complete LDA+G method with full
charge self-consistency and without tight-binding fitting.  Two
important factors have to be considered for this purpose: (1)
Realistic materials consist of non-strongly-correlated bands as well,
which can be treated nicely by LDA, and those strongly-correlated
bands, which require the Gutzwiller step. Proper separation of two
sets of energy bands is therefore necessary; (2) Full charge density
self-consistency need to be considered. These will be the main points
for our following discussions.

We first divide the complete orbital basis into localized and extended
orbitals, and the interactions are added only for localized orbitals,
for example $d$ or $f$ orbitals in transition metal or rare earth
compounds .  The localized and extended orbitals are labeled by
\{$\vert i\sigma\rangle=C_{i\sigma}^{\dagger}\vert0\rangle$\} and
$\{\vert i\delta\rangle=C_{i\delta}^{\dagger}\vert0\rangle\}$
respectively, and the completeness of orbital basis requires that:
\begin{equation}
  \sum_{i\sigma}\vert i\sigma\rangle\langle i\sigma\vert+\sum_{i\delta}\vert i\delta\rangle\langle i\delta\vert=1\label{eq:localcompleteness}
\end{equation}
Under the representation of this basis set, the $H_{LDA}$ reads
\begin{equation}
\begin{split} 
  H_{LDA}
  & =(\sum_{i\sigma}\vert i\sigma\rangle\langle i\sigma\vert+\sum_{i\delta}\vert i\delta\rangle\langle i\delta\vert)\\
  & \quad H_{LDA}(\sum_{j\sigma'}\vert j\sigma'\rangle\langle
  j\sigma'\vert+\sum_{j\delta'}\vert j\delta'\rangle\langle
  j\delta'\vert)
\end{split}
\end{equation}

As discussed in Section I, the GWF $\vert\Psi_{G}\rangle$ is
constructed from the HWF $\vert\Psi_{0}\rangle$ with proper
projection. Any operator acting on GWF can be mapped to an Gutzwiller
effective operator which acts on HWF instead.  Since $H_{LDA}$ only
consists of single particle operators, following the definition in
Eq.~(\ref{definition_G}), its corresponding Gutzwiller effective
Hamiltonian can be written as,
 \begin{equation}
\begin{split} 
 H_{LDA}^{G}
 & =(\sum_{i\sigma}z_{i,\sigma}\vert i\sigma\rangle\langle i\sigma\vert+\sum_{i\delta}\vert i\delta\rangle\langle i\delta\vert)\\
 & \quad H_{LDA}(\sum_{j\sigma'}z_{j,\sigma'}\vert j\sigma'\rangle\langle j\sigma'\vert+\sum_{j\delta'}\vert j\delta'\rangle\langle j\delta'\vert)\\
 & +\sum_{i\sigma}(1-z_{i\sigma}^{2})\vert i\sigma\rangle\langle i\sigma\vert H_{LDA}\vert i\sigma\rangle\langle i\sigma\vert\end{split}
\end{equation}
To derive this Gutzwiller effective Hamiltonian, it is essential to
understand that for those non-interacting orbitals, the corresponding
renormalization fact $z_{i\delta}$ is equal to 1. This formula could
be further simplified using the completeness
condition~(\ref{eq:localcompleteness})
\begin{equation}
\begin{split} 
  H_{LDA}^{G}
  & =(\sum_{i\sigma}z_{i,\sigma}\vert i\sigma\rangle\langle i\sigma\vert+1-\sum_{i\sigma}\vert i\sigma\rangle\langle i\sigma\vert)\\
  & \quad H_{LDA}(\sum_{j\sigma'}z_{j,\sigma'}\vert j\sigma'\rangle\langle j\sigma'\vert+1-\sum_{j\sigma'}\vert j\sigma'\rangle\langle j\sigma'\vert)\\
  & +\sum_{i\sigma}(1-z_{i\sigma}^{2})\vert i\sigma\rangle\langle
  i\sigma\vert H_{LDA}\vert i\sigma\rangle\langle
  i\sigma\vert
\end{split}
\end{equation}
and the interaction energy is given as,
\begin{equation}
  \langle\Psi_{G}\vert H_{int}\vert\Psi_{G}\rangle=\sum_{i;\Gamma}E_{i\Gamma}m_{i\Gamma}\label{eq:int}
\end{equation}

Now it is clear that the complete basis set defined at beginning is
actually not necessary for realistic calculations, because only the
 localized orbitals $\langle i\sigma\vert$ appear in the above
equation. The interaction term are also defined only for those
localized orbitals. We then come to a stage very similar to LDA+U,
where localized orbitals are defined and interaction within those
orbitals is supplemented. What is in additional to LDA+U scheme is that
the kinetic energy of each local orbitals is renormalized by factor
$z_{i\sigma}$ which need to be determined in terms of configuration
weights and configuration energy through the variational approach as
shown below.

In realistic calculations for solid crystals, it is more convenient to
carry out the calculations in reciprocal space, especially for those
plane wave methods. The transformation to the reciprocal space is
quite straightforward, because the Gutzwiller approximation keeps the
translational symmetry unbroken.  We first define the Bloch states of
localized orbitals $\vert i\sigma\rangle$
\begin{equation}
  \vert{k\sigma}\rangle=\frac{1}{N}\sum_{i}e^{ikR_{i}}\vert
  i\sigma\rangle\label{eq:localorbitals_k}
\end{equation} 
Then the Gutzwiller effective Hamiltonian $H_{LDA}^{G}$ in $k$-space
can be written as,
 \begin{equation}
\begin{split}  H_{LDA}^{G}
 & =(\sum_{k\sigma}z_{\sigma}\vert k\sigma\rangle\langle k\sigma\vert+1-\sum_{k\sigma}\vert k\sigma\rangle\langle k\sigma\vert)\\
 & \quad H_{LDA}(\sum_{k'\sigma'}z_{\sigma'}\vert k'\sigma'\rangle\langle k'\sigma'\vert+1-\sum_{k'\sigma'}\vert k'\sigma'\rangle\langle k'\sigma'\vert)\\
 & +\sum_{kk'\sigma}(1-z_{\sigma}^{2})\vert k\sigma\rangle\langle k'\sigma\vert H_{LDA}\vert k'\sigma\rangle\langle k\sigma\vert\end{split}
\end{equation}
Let's define the projector
$\hat{P}=\sum_{k\sigma}\hat{P}_{k\sigma}=\sum_{k,\sigma}\vert
k\sigma\rangle\langle k\sigma\vert$ which projects onto the Bloch
state of localized orbital, then the projection to the remaining
delocalized orbitals is taken into account by $1-\hat{P}$. For
convenience, here we also define another projector
$\hat{Q}=\sum_{k,\sigma}z_{\sigma}\vert k\sigma\rangle\langle
k\sigma\vert$. Then we have
 \begin{equation}
\label{eq:HLDAG_ex}
\begin{split}  H_{LDA}^{G}
 & =(1-\hat{P}+\hat{Q})H_{LDA}(1-\hat{P}+\hat{Q})\\
 & +\sum_{kk'\sigma}(1-z_{\sigma}^{2})\vert k\sigma\rangle\langle k'\sigma\vert H_{LDA}\vert k'\sigma\rangle\langle k\sigma\vert
\end{split}
\end{equation}

The total energy is obtained by evaluating the expectation value of
the Hamiltonian,
\begin{equation}
\label{eq:totalenergy_G}
\begin{split}
 E(\rho) 
 & =\langle\Psi_{0}\vert H_{LDA}^{G}\vert\Psi_{0}\rangle+\sum_{\Gamma}E_{\Gamma}m_{\Gamma}-E_{dc}\\
 & =\langle\Psi_{0}\vert(1-\hat{P}+\hat{Q})H_{LDA}(1-\hat{P}+\hat{Q})\vert\Psi_{0}\rangle\\
 &
 +\sum_{\sigma}(1-z_{\sigma}^{2})n_{\sigma}\mathcal{E}_{LDA}^\sigma+\sum_{\Gamma}E_{\Gamma}m_{\Gamma}-E_{dc}
\end{split}
\end{equation}
in which $\mathcal{E}_{LDA}^\sigma=\sum_{k}\langle k\sigma\vert
H_{LDA}\vert k\sigma\rangle$, $n_{\sigma}=\sum_{k}\langle\Psi_{0}\vert
k\sigma\rangle\langle k\sigma\vert\Psi_{0}\rangle$, and $E_{dc}$ is
the double counting energy.

Now the remaining task is to minimize the total energy functional with
respect to variational parameters: uncorrelated wave function
$|\Psi_{0}\rangle$ and atomic configuration weight $m_{\Gamma}$. Very
similar to the familiar Kohn-Sham equation, the uncorrelated wave
function $|\Psi_{0}\rangle$ in the periodic lattice can be written as
a simple Slater Determinant of single particle wave functions
$|\psi_{nk}\rangle$.  Two sets of variational equations can be derived
from minimization of Eq.~(\ref{eq:totalenergy_G}) with respect to
$|\psi_{nk}\rangle$ and $m_{\Gamma}$, respectively. For the variation
with respect to $|\psi_{nk}\rangle$, please note that the orbital
occupation number $n_\sigma=\sum_{nk}\langle
\psi_{nk}|\hat{P}_{k\sigma}|\psi_{nk}\rangle$ also depends on
$|\psi_{nk}\rangle$ inexplicitly.
\begin{equation}
\begin{split}\frac{\partial E(\rho)}{\partial f_{nk}\langle\psi_{nk}|}= & [H_{LDA}^{G}+\frac{\partial E(\rho)}{\partial z_{\sigma}}\frac{\partial z_{\sigma}}{\partial\; n_{\sigma}}\hat{P}_{k\sigma}-H_{dc}]|\psi_{nk}\rangle\\
= & \epsilon_{nk}|\psi_{nk}\rangle\end{split}
\label{eq:Evar_psi_nk}\end{equation}
\begin{equation}
\frac{\partial E(\rho)}{\partial m_{\Gamma}}=\sum_{\sigma}\frac{\partial E(\rho)}{\partial z_{\sigma}}\frac{\partial z_{\sigma}}{\partial\; m_{\Gamma}}+E_{\Gamma}=0\label{eq:Evar_m_nk}\end{equation}
in which 
\begin{equation}
\begin{split}\frac{\partial E(\rho)}{\partial z_{\sigma}}= & \frac{2}{z_{\sigma}}(\frac{1}{2}\sum_{nk}f_{nk}\langle\psi_{nk}|\hat{P}_{k\sigma}H_{LDA}^{G}+H_{LDA}^{G}\hat{P}_{k\sigma}|\psi_{nk}\rangle\\
 & -n_{\sigma}\mathcal{E}_{LDA}^\sigma)\end{split}
\label{eq:Evar_psi_nk_2}\end{equation}
 When deriving this equation, the following relation is used
 \begin{equation}
\begin{split} & z_{\sigma}\frac{\partial(1-\hat{P}+\hat{Q}))}{\partial z_{\sigma}}\\
= & z_{\sigma}\vert k\sigma\rangle\langle k\sigma\vert\\
= & (1-\hat{P}+\hat{Q})\vert k\sigma\rangle\langle k\sigma\vert\\
= & (1-\hat{P}+\hat{Q})\hat{P}_{k\sigma}\end{split}
\end{equation}
There are several constraints. The wave functions should be orthogonal and normalized, the total
configurations weight must be unity, and for pure density correlations
the local densities will not be changed in GVA:
\begin{equation}
\begin{split}\langle\psi_{nk}|\psi_{n^{\prime}k^{\prime}}\rangle & =\delta_{n,n^{\prime}}\delta_{k,k^{\prime}}\\
  \sum_{\Gamma}m_{\Gamma} & =1 \\
  \sum_{\Gamma }\left\langle \Gamma \right\vert C_{\sigma }^{\dag }C_{\sigma
  }\left\vert \Gamma \right\rangle m_{\Gamma }&=\langle\hat{n}_{\sigma}\rangle_{G}
  =\langle\hat{n}_\sigma\rangle_0
\end{split}
\label{eq:constraints}\end{equation}
Through the above steps, we will be able to solve the problem for
fixed LDA Hamiltonian $H_{LDA}$.

Now the question is how can we achieve self-consistency in the charge
density. This step is very crucial, and the reason is the following.
As we discussed above, all electrons (both delocalized and localized)
should be included in realistic calculations.  However those
delocalized orbitals are treated in LDA level, and localized states
are treated by the LDA+G step. The modification of the localized state
will in return affect the charge distribution of all other delocalized
state, particularly the charge transfer process between the
delocalized and the localized orbitals may happen. This is of course
important physics. If it is not treated properly, different
conclusions may be drawn, as already discussed in the example studies
for Na$_{1-x}$CoO$_2$~\cite{NaCoO}, where several post-LDA plus DMFT
studies give different results.

The charge density self-consistency can be achieved easily as long as
the charge density can be constructed, because the LDA Hamiltonian
is determined by the electron density. In the present LDA+G scheme,
the electron density can be constructed from the Gutzwiller wave
functions by:
\begin{equation}
\rho=\langle\Psi_{G}|\hat{\rho}|\Psi_{G}\rangle=\langle\Psi_{0}|\hat{\rho^{G}}|\Psi_{0}\rangle
\end{equation}
Since $\hat{\rho}$ is also a one-particle operator, similar to
previous steps, with the help of Eq.~(\ref{eq:oneparticleoperator}) we
have
 \begin{equation}
\begin{split}  \hat{\rho}^{G}
 & =(\sum_{i\sigma}z_{i,\sigma}\vert i\sigma\rangle\langle i\sigma\vert+1-\sum_{i\sigma}\vert i\sigma\rangle\langle i\sigma\vert)\\
 & \quad|r\rangle\langle r|(\sum_{j\sigma'}z_{j,\sigma'}\vert j\sigma'\rangle\langle j\sigma'\vert+1-\sum_{j\sigma'}\vert j\sigma'\rangle\langle j\sigma'\vert)\\
 & +\sum_{i\sigma}(1-z_{i\sigma}^{2})\vert i\sigma\rangle\langle i\sigma\vert|r\rangle\langle r|\vert i\sigma\rangle\langle i\sigma\vert\end{split}
\end{equation}
or we can write down the expression in the momentum space with the
following simple expression,
\begin{equation}
\label{eq:density_oper_G}
\begin{split}  \hat{\rho}^{G}
 & =(1-\hat{P}+\hat{Q})|r\rangle\langle r|(1-\hat{P}+\hat{Q})\\
 & +\sum_{kk'\sigma}(1-z_{\sigma}^{2})\vert k\sigma\rangle\langle k'\sigma\vert|r\rangle\langle r|\vert k'\sigma\rangle\langle k\sigma\vert
\end{split}
\end{equation}

Eq.~(\ref{eq:Evar_psi_nk}), (\ref{eq:Evar_m_nk}) together with
Eq.~(\ref{eq:zfac}) and (\ref{eq:density_oper_G}) provide a
self-consistent scheme, which is named LDA+Gutzwiller method by
us. Eq.~(\ref{eq:Evar_psi_nk}) is similar to the KS-equation in LDA or
GGA, except that the Hamiltonian has been replaced by the
corresponding effective Gutzwiller one with orbital-dependent terms.
Eq.~(\ref{eq:Evar_m_nk}), which is used to determine the configuration
weight, and Eq.~(\ref{eq:zfac}), which determine the factors
$z_{\sigma}$, are newly introduced by GVA.

\begin{figure}[tbp]
\includegraphics[trim=15mm 145mm 110mm 15mm,scale=0.9]{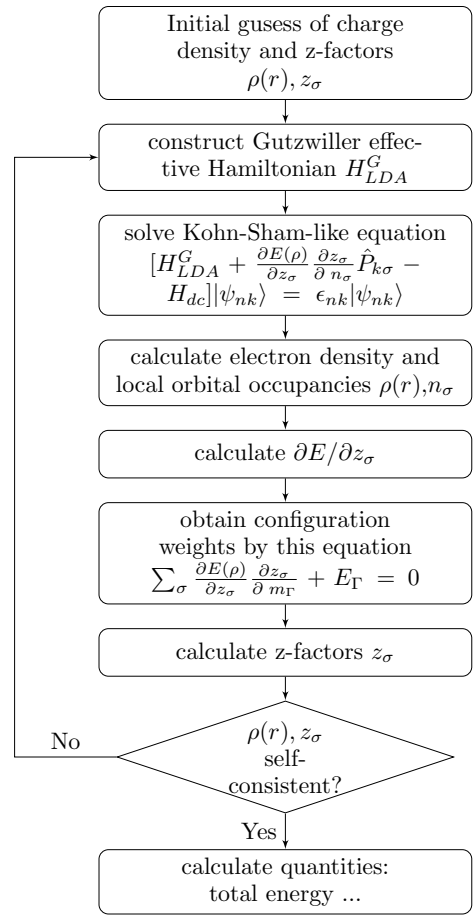}
\caption{Flow chart of self-consistent loops for LDA+Gutzwiller
  method.}
\label{fig:flowchartsforldag}
\end{figure}

We illustrate our schematic self-consistent loops for LDA+G method in
Fig.~\ref{fig:flowchartsforldag}. Two main steps in this scheme are:
(1) for fixed factor $z_{\sigma}$, solving the Kohn-Sham-like equation
to get uncorrelated wave function $\Psi_{0}$. (2) for fixed
$\Psi_{0}$, calculate the configuration weights $m_{\Gamma}$ and then
obtain factor $z_{\sigma}$. The iteration loops end when both electron
density and renormalization factors are self-consistent. This scheme
could be easily implemented in all kinds of existing \textit{ab
  initio} codes, no matter what kinds of basis set are used for the
wave-function, because the essential computational requirement is just
the calculations of projection to some local orbitals. If the LDA+U
method is already available in the original code, the implementation
of LDA+G method will be much easier, since the local orbitals defined
in LDA+U can be also used in the LDA+G formalism.

After the charge density self consistency has been achieved, the
ground state properties such as the stable crystal structure, magnetic
structure as well as the elastic properties can be calculated, which
is quite similar with the standard LDA procedure. Besides that, we can
also obtain the density of states by LDA+G.  Because the band
dispersion $E_{nk}$ obtained by LDA+G is for the quasi-particle
excitations, we can derive two types of density of states (DOS) from
the LDA+G band structure. The first one is the quasi-particle DOS,
which can be expressed as
\begin{equation}
\rho_{QP}\left(\omega\right)=\sum_{nk}\delta\left(\omega-E_{nk}\right)
\end{equation}
Experimentally the electronic part of the low temperature specific
heat is directly determined by the quasi-particle DOS and thus can be
estimated by LDA+G for correlated materials. Another type of DOS is
the integrated electronic spectral function, which is called electron
DOS in LDA+G. Using $z_{\sigma}$ obtained by the Gutzwiller
approximation, the weight of the quasi-particle peak at $E_{nk}$
appearing in the electronic spectral function can be expressed as
\begin{equation}
\mathit{\mathcal{Z_{\mathrm{nk}}=}}\left\langle
  kn\left|\hat{Q}^{2}\right|nk\right\rangle +\left\langle
  kn\left|1-\hat{P}\right|nk\right\rangle 
\end{equation}
Then the electron DOS can be
obtained by the summation over all $k$ ,which reads
\begin{equation}
\rho_{el}\left(\omega\right)=\sum_{nk}\mathcal{Z}_{\mathrm{nk}}\delta\left(\omega-E_{nk}\right)\end{equation}

As we mentioned in the previous section, only the coherent parts(quasi-particle
peaks) can be captured by LDA+G not the incoherent parts (Hubbard bands).
The electron DOS in LDA+G corresponds to
the low energy part of the photo emission spectrum, which is mainly
determined by the quasi-particle dynamics.

In the end, we would like to comment on the relationship between LDA+G
method and LDA or LDA+U.  First it can be easily figured out that,
LDA+G method falls back to DFT-LDA method spontaneously in
non-correlated systems.  This can be seen from the multi-band Hubbard
model in which on-site interactions go to zero, then no configuration
is energetically unfavorable and GWF falls back to HWF. In LDA+G
method, this case means that the supplemented on-site interaction
energy and double-counting term are zero, then all localized orbitals
are really delocalized and the corresponding renormalization
parameters are $Z_{\sigma}=1$. The formalism falls back exactly to
DFT-LDA.

LDA+Gutzwiller method can also cover the LDA+U method when applying to
strongly correlated insulators with long range ordering, in which LDA
fails while LDA+U makes its success. Actually this is quite easy to be
understood. In the strongly correlated insulator with long range
ordering and integer occupation, such as the anti-ferromagnetic phase
in the half filled Hubbard model, the unit cell is doubled by the AF
order, which greatly reduces the local fluctuation among the atomic
configurations and thus increases the z-factor to be close to
unity. Again in the $z$=1 limit, the GWF returns back to HWF, the
LDA+G energy functional is equivalent to that of LDA+U when the
$z$-factors approach one.

\subsection{Derivation from Gutzwiller Density Functional Theory}

In the last section, we have derived the LDA+G method in a physical
but yet not rigorous way. In this section, however, we will derive the
LDA+G method from a sound base. We will first introduce a Gutzwiller
density functional theory (GDFT), which is rigorous and exact, just
like the Kohn-Sham (KS) formalism developed from the density
functional theory (DFT). In the KS formalism, as long as we know the
functional of exchange-correlation energy $E_{xc}^{KS}$, we can solve
the ground state problem. This is also true for GDFT, where we
have $E_{xc}^G$ in its rigorous form instead of $E_{xc}^{KS}$. Of
course, exact $E_{xc}^G$ is unknown, and certain kinds of
approximation have to be used in realistic calculations. In the KS
formalism, if the LDA is used for the $E_{xc}^{KS}$ ($\approx
E_{xc}^{LDA}$), then the LDA-KS type formalism is realized; in
addition, if LDA+U approximation is used for the $E_{xc}^{KS}$
($\approx E_{xc}^{LDA+U}$), the LDA+U method can be obtained. We will
show here that the LDA+G method can be actually regarded as a ``{\it
  LDA+U approximation in GDFT formalism}'', where LDA+U type
approximation is used for the exchange correlation term.  We will
first establish the GDFT, then we derive the LDA+G formalism from the
firm base of GDFT.

\subsubsection{DFT and Kohn-Sham}

It is helpful to recall the basis of DFT first. The Hohenberg-Kohn
(HK) theorem\citet{DFT_HK} shows that the total energy of a
interacting electron system can be defined as a universal functional
in terms of electron density $\rho(r)$.  The ground state energy is
the global minimum of the functional. The electron density that
minimizes the functional is the exact ground state electron
density. The equations are written as
\begin{eqnarray}
E[\rho] &=& \langle\Psi\vert H\vert\Psi\rangle = T[\rho]+E_{int}[\rho]  \nonumber \\
\rho &=&\langle\Psi\vert \hat{\rho} \vert\Psi\rangle
\label{eq:HK}
\end{eqnarray}
where $\vert\Psi\rangle$ is the ground state many-body wave function,
$T$ is the kinetic energy, and $E_{int}$ is the interaction energy. At
this stage for simplicity, the energy due to external potential are
not included in the formula (it will be supplemented later). This
theorem is exact, but can not be used directly since the explicit form
of this functional is unknown. This problem can be transferred to a
equivalent Kohn-Sham (KS) problem by using the well-known Kohn-Sham
\textit{ansatz}\citet{DFT_KS}, which is now become one of the most
important basis of first
principle electronic structure calculations for solid states. In this \textit{ansatz}, a
reference system is introduced, whose exact Hamiltonian is still
unknown, but we know that its ground state wave function can be
exactly written as the HWF $|\Psi_0\rangle$. (Therefore, the reference
system here is actually a non-interacting system because we know that
its wave function is $|\Psi_0\rangle$). As long as the charge density
of the reference system $\rho^0$ matches the true ground state charge
density (i.e, $\rho^0=\rho$), then from the Hohenberg-Kohn DFT, the
total energy of the true system can be reproduced through the
reference system,
\begin{equation}
\begin{split}
  E[\rho]=&E^{KS}[\rho^0]=T^{KS}[\rho^0]+E_H^{KS}[\rho^0]+E_{xc}^{KS}[\rho^0] \\
         =&\langle\Psi_0|\hat{T}|\Psi_0\rangle +E_H^{KS}[\rho^0]+E_{xc}^{KS}[\rho^0] \\
  \rho=&\rho^0=\langle\Psi_0\vert \hat{\rho} \vert \Psi_0\rangle
\end{split}
\end{equation}

The KS kinetic energy $T^{KS}$ and the Hartree energy $E_H^{KS}$ of
the reference system is different with the true kinetic and
interaction energy, $T$ and $E_{int}$, but importantly their
functional forms are known. KS's idea is simply to re-organize the
total energy expression, such that those known parts can be treated
explicitly and all unknown parts are moved into the third term called
exchange-correlation energy $E_{xc}^{KS}$. It is in this sense that KS
formalism is still exact for the ground state total energy. The merit
of KS formalism is that the problem is solvable as long as the
functional form $E_{xc}^{KS}$ is known. By definition, the
exchange-correlation energy is given as,
\begin{equation}
  E_{xc}^{KS}=\Delta T^{KS}+ \Delta E_{int}^{KS}=(T-T^{KS})+(E_{int}-E_H^{KS})
\end{equation}
where two contributions should be physically included: (1) the
correction to the kinetic energy $\Delta T^{KS}$; (2) the correction
to the interaction energy $\Delta E_{int}^{KS}$.

The KS formalism up to now is exact, however certain kinds of
approximations have to be made for $E_{xc}^{KS}$ in realistic
calculations, as will be discussed below. Nevertheless, here we want
to put a note for the reference system. Once the approximation has
been introduced for the exchange-correlation term $E_{xc}^{KS}$, the
nature of the reference system may be modified. In reality, if the
LDA+U approximation is used for the exchange-correlation potential in
KS formalism, the reference system is no-longer non-interacting, and
the $|\Psi_0\rangle$ is just a approximate wave function.

\subsubsection{Gutzwiller Density Functional Theory (GDFT)}

Here we can establish a exact Gutzwiller density functional theory
(GDFT) in parallel to KS formalism. One can see from KS formalism
that, taking a non-interacting system as reference is not a
necessity. What is really important is to know the exact form of the
wave-function of the reference system, such as $|\Psi_0\rangle$ in KS.
The main benefit of this strategy is that in the reference system the
kinetic operator can be evaluated explicitly in quite a simple
form. In the spirit of Kohn-Sham \textit{ansatz}, any systems can be
taken as reference providing that it has the same electron density as
the true system. We can theoretically formulate the exact Gutzwiller
density functional theory (GDFT), similar to KS, as follows.

(1) To replace the original difficult interacting many-body system, we
choose a auxiliary reference system, whose exact Hamiltonian is still
unknown, but we know that its ground state wave function is given as
the Gutzwiller wave function $|\Psi_G\rangle$ (rather than the HFW
$|\Psi_0\rangle$).  An important point to be noticed here is that the
fact whether the reference system is interacting or non-interacting
actually does not matter, since the Gutzwiller wave function can be
used to describe both the interacting and non-interacting systems in a
better way. (The nature of the reference system depends on the choice
of exchange-correlation potential as will be further discussed in the
following part); (2) Following the KS {\it ansatz}, we assume that the
ground state density of the original interacting system is equal to
that of the reference system $\rho^G=\rho$. (The representability is
not rigorously proofed at this stage, therefore this step is still an
{\it ansatz}. However, considering the fact that the $\Psi^G$
automatically return back to $\Psi_0$ in the non-interacting limit, it
has the same spirit as KS {\it ansatz} which has been proofed to be
valid for many applications.) (3) The kinetic energy of the reference
system can be explicitly written as $T^G=\langle \Psi_G\vert \hat{T}
\vert \Psi_G\rangle$. (4) All unknown parts are moved into the
exchange-correlation energy $E_{xc}^G$. Finally, the total energy and
the charge density will be written as,
\begin{equation}
\begin{split}
  E[\rho]=&E^G[\rho^G]=T^G[\rho^G]+E_H^G[\rho^G]+E_{xc}^G[\rho^G] \\
         =&\langle\Psi_G|\hat{T}|\Psi_G\rangle+E_H^G[\rho^G]+E_{xc}^G[\rho^G] \\
  \rho=&\rho^G=\langle \Psi^G \vert \hat{\rho} \vert \Psi^G \rangle
\end{split}
\end{equation}

The above formulation of GDFT is still exact similar to KS. We can
also come to the same conclusion as KS that if the exact
exchange-correlation energy $E_{xc}^G$ is known, the exact ground
state energy of the true system will be obtained.  Again for the
physical understanding, two terms are included in the
exchange-correlation energy: (1) the correction to the kinetic energy
$\Delta T^G$; (2) the correction to the interaction energy $\Delta
E_{int}^G$, as expressed as,
\begin{equation}
  E_{xc}^{G}=\Delta T^G+ \Delta E_{int}^G=(T-T^{G})+(E_{int}-E_H^{G})
\end{equation}

As already mentioned, the $\Psi_G$ automatically return back to
$\Psi_0$ in the non-interacting limit, therefore the present GDFT can
be regarded as a general extension of original KS formalism.

\subsubsection{Approximations for $E_{xc}$}

Certain approximation has to be introduced for the unknown
exchange-correlation part in order to perform practically
calculations. We will start from the LDA and LDA+U approximations used
in KS formalism, and then we will show that by introducing the LDA+U
type approximation in GDFT, the LDA+G method can be derived.

(1) {\it LDA or GGA}

The most popularly used and widely accepted approximation is the LDA,
where the exchange-correlation energy is approximated as,
\begin{equation}
E_{xc}^{KS}\approx E_{xc}^{LDA}=\Delta T^{LDA}+\Delta E_{int}^{LDA}
\end{equation}
We assume the readers have the basic knowledge about LDA, we therefore
do not discuss its details here. The only point we want to emphasize
is that the LDA is basically parametrized from the uniform electron
gas, and only the {\it local} part of the exchange-correlation
potential is kept, in other words, the {\it non-local} part of the
exchange-correlation potential is neglected. This is the reason why
LDA works well for simple metals, such as the Na, K, where wide $s$
band cross the Fermi level, but it fails for strongly correlated
systems, such as transition-metal oxides.

(2){\it LDA+U for strongly correlated systems}

To overcome the problem of LDA for strongly correlated systems, LDA+U
method has been introduced, where the exchange-correlation energy is
approximated as,
\begin{equation}
\begin{split}
E_{xc}^{KS}\approx & E_{xc}^{LDA+U}=\Delta T^{LDA}+\Delta E_{int}^{LDA+U} \\
=&\Delta T^{LDA}+\Delta E_{int}^{LDA}+\langle H_{int} \rangle_0-E_{dc} \\
=&E_{xc}^{LDA}+\langle H_{int} \rangle_0-E_{dc}
\end{split}
\end{equation}
The spirit of LDA+U method is that the LDA do not treat the
interaction energy sufficiently well, and it need to be corrected for
strongly correlated systems. Therefore, the interaction energy
correction in original LDA $\Delta E_{int}^{LDA}$ is replaced by the
LDA+U counterpart $\Delta E_{int}^{LDA+U}=\Delta E_{int}^{LDA}+\langle
H_{int} \rangle_0-E_{dc}$. In such a way, the interaction term is
treated more explicitly, and the energy is improved.  

To understand the LDA+U formalism, it is very important to notice the
following points: (1) the reference system (in KS) is no longer
non-interacting anymore, since the interaction term $H_{int}$ is
supplemented from the exchange-correlation potential. Therefore, the
$|\Psi_0\rangle$ is no longer the rigorous eigen state of the
reference system, instead it is just a approximation; (2) It is only
up to this step that the definition of local orbital and interaction
strength $U$ is necessary. In such a way, the orbital-dependent
potential is introduced (the LDA type $E_{xc}^{LDA}$ only depends on
the density, not on the orbital).

The LDA+U method is quite successful for many of the insulating
systems, which have AF long-range ordered ground state, however its
quality is still not sufficient and can be improved further. The main
drawbacks of LDA+U are two folds: (1) The supplemented interacting
term $H_{int}$ is treated by crude Hartree-scheme, and the interaction
energy is typically overestimated. Actually, once the reference system
becomes interacting system, the $\Psi_0$ is no-longer a rigorous wave
function; (2) Due to the usage of $\Psi_0$ as an approximate wave
function, only the interaction energy is further corrected (over LDA),
namely $\langle H_{int} \rangle_0$ is the further correction to the
interaction energy. But the kinetic part $\Delta T$ is still kept to
be the same as that in LDA ($\Delta T^{LDA}$). However it is known
that the presence of interaction term should also renormalize the
kinetic energy. Those drawbacks can be improved from our LDA+G
formalism as will be derived below.

\subsubsection{Derivation of LDA+G from the GDFT}

The GDFT itself is also exact, however, certain approximations have to
be used for the $E_{xc}^G$ in practical calculations.  Since the
exchange-correlation energy is a functional of charge density, the
easiest way of course is still to use the local density approximation,
and neglect the non-local part of the potential, i.e, $E_{xc}^G\approx
E_{xc}^{LDA}$. For the Hartree energy, it only depends on the charge
density, and it is the same for both KS-DFT and GDFT,
i.e. $E_{H}^G=E_{H}^{KS}$. Therefore, after applying the LDA to
$E_{xc}^G$, all the potential energies in GDFT recover to be the same
as that in LDA-KS. In this limit, we already know that the reference
system is a non-interacting system, and the wave function $\Psi_G$
should return back to $\Psi_0$. Therefore, all the above GDFT
formalism returns back to LDA-KS if the LDA is used for $E_{xc}^G$. So
far, we gain nothing from the usage of GDFT. However, if the LDA+U
type approximation is used for $E_{xc}^G$ in GDFT, the situation will
be much improved.

As have been discussed above, to overcome the problem of LDA for
strongly correlated system, the strategy of LDA+U approximation is to
introduce a supplemented interaction term $H_{int}$ in the exchange
correlation potential, such that the electron-electron interaction can
be treated more explicitly beyond LDA. The reference system now is no
longer non-interacting, but $\Psi_0$ is still used to approximate the
wave function of the reference system in the LDA+U KS formalism. The
same approximation for the exchange correlation term can be also used
in the GDFT, namely a interaction term $H_{int}$ can be supplemented
in the exchange correlation potential to describe the electrons in localized
orbitals better (beyond uniform electron gas in LDA). Again, the
reference system is a interacting system now, however, what is
different in GDFT is that the wave function of reference system is
given as $\Psi_G$ rather than $\Psi_0$. Of course, the GWF $\Psi_G$ is
much better than HWF $\Psi_0$ for interacting system, and it is in
this sense that the formalism is improved over LDA+U.

Therefore, using the similar LDA+U approximation in the GDFT,
exchange-correlation energy and the corresponding total energy can be
written as,
\begin{equation}
\begin{split}  
E[\rho]=&T^G[\rho]+E_H[\rho]+E_{xc}^G[\rho] \\
       =&\langle\Psi_G|\hat{T}|\Psi_G\rangle +E_H[\rho]+E_{xc}^G[\rho] \\
E_{xc}^{G}\approx & E_{xc}^{LDA+G}=\Delta T^{LDA}+\Delta E_{int}^{LDA+G} \\
=&\Delta T^{LDA}+\Delta E_{int}^{LDA}+\langle H_{int} \rangle_G-E_{dc} \\
=& E_{xc}^{LDA}+\langle H_{int} \rangle_G-E_{dc}
\end{split}
\label{LDA+G}
\end{equation}
It differs from the LDA+U KS scheme in the following two points: (1)
The supplemented interaction term is more precisely dealt with the
Gutzwiller wave function, namely the $\langle H_{int} \rangle_G$ is
used instead of $\langle H_{int} \rangle_0$; (2) Although the $\Delta
T^{LDA}$ is still used in the exchange-correlation functional, the
kinetic energy is actually improved through the replacement of
$T^{KS}$ by $T^G$.  The usage of $\Delta T^{LDA}$ in $E_{xc}^G$
requires more discussions. It is known that the drawback of $\Delta
T^{LDA}$ is that it keeps only the local part and neglects the
non-local part. Therefore, to make improvement, non-local correction
should be supplemented. However, it is seen from our above GDFT
formalism, the non-local correction to the kinetic energy has been
naturally included through the replacement of
$\langle\Psi_0|\hat{T}|\Psi_0\rangle$ by
$\langle\Psi_G|\hat{T}|\Psi_G\rangle$. Therefore, in the
exchange-correlation part, the non-local correction to the kinetic
energy is no-longer necessary, and only local-part need to be
considered. This is the reason why $\Delta T^{LDA}$ can be used for
$E_{xc}^G$. The usage of $\Delta T^{LDA}$ in $E_{xc}^G$ can also
guaranty that the present LDA+G formalism return back to the LDA+U
solution in the static limit, where the $z$-factors approaching unity
and $|\Psi^G\rangle$ approaching $|\Psi^0\rangle$.

Up to this stage, we have finished all the necessary steps, and we
show that the original LDA+G formalism discussed in the last section
can be derived from a more rigorous base, namely the ``LDA+U type
approximation for the exchange correlation potential in the exact GDFT
formalism''. Using the Eq.~(\ref{LDA+G}), the Hamiltonian~(\ref{LDA++})
discussed in the last section under GWF will be recovered. To be more
practical, here we will write down the final version of the necessary
equations explicitly. The total energy and the exchange-correlations
read,
\begin{equation}
\begin{split} 
  E[\rho]=&\langle\Psi_{G}\vert
  \hat{T}\vert\Psi_{G}\rangle+E_{H}[\rho]+\int\!\! V_{ext}\,\rho
  d^{3}r+E_{xc}^{G}[\rho] \\
  E_{xc}^G\approx & E_{xc}^{LDA+G}=E_{xc}^{LDA}+\langle\Psi_G\vert H_{int}
  \vert\Psi_G\rangle-E_{dc} \\
%  \rho(r)=&\langle\Psi_{G}\vert \hat{\rho}
%  \vert\Psi_{G}\rangle=\langle\Psi_{0}\vert\hat{\rho}^{G}\vert\Psi_{0}\rangle
\end{split}
\label{eq:Etot_rho_G}
\end{equation}
in which the electron density now is,
 \begin{equation}
\begin{split}\rho(r)=\langle\Psi_{G}\vert r\rangle\langle r\vert\Psi_{G}\rangle=\langle\Psi_{0}\vert\hat{\rho}^{G}\vert\Psi_{0}\rangle\end{split}
\label{eq:density_g}\end{equation}
and the kinetic energy operator is
\begin{equation}
T=\sum_{i}-\frac{1}{2}\nabla_{i}^{2}
\end{equation}
where $i$ is electron label.  The Hartree
interaction energy of electrons is
\begin{equation}
  E_{H}=\frac{1}{2}\int\! d^{3}rd^{3}r^{\prime}\frac{\rho(r)\rho(r^{\prime})}{|r-r^{\prime}|}
\end{equation}

If the supplemented interaction part of the Hamiltonian is diagonal in
configuration space (This is true if only density-density interactions
are considered), the interaction term in the exchange-correlation
energy reads according to Eq.~ (\ref{eq:H_at}), (\ref{eq:UiI}),
(\ref{eq:mgutzH}),
\begin{equation}
  \langle\Psi_G\vert H_{int} \vert \Psi_G\rangle
  =\langle\Psi_{G}\vert\sum_{i}H_{i}\vert\Psi_{G}\rangle=\sum_{i;\Gamma}E_{i\Gamma}m_{i\Gamma}
\label{eq:Eint}
\end{equation}

For convenience, external potential energy, Hartree energy and
exchange-correlation energy (of LDA part) can be grouped together, and
written as an functional form of charge density,
 \begin{equation}
   E_{eHxc}[\rho]=E_{H}[\rho]+\int\! V_{ext}\;\rho\; d^{3}r+E_{xc}^{LDA}[\rho]
\label{eq:EeHxc}
\end{equation}
and the effective potential is defined as
 \begin{equation}
   V_{eHxc}=\frac{\delta E_{eHxc}[\rho]}{\delta\rho}
\label{eq:V_eHxc}
\end{equation}
which is exactly the same as that in the LDA-KS formalism.

Now the total energy is,
\begin{equation}
\begin{split}
  E[\rho,m_{i\Gamma}]=& \langle\Psi_{G}\vert T\vert\Psi_{G}\rangle+E_{eHxc}[\rho]+\sum_{i;\Gamma}E_{i\Gamma}m_{i\Gamma}-E_{dc}\\
 =& \langle\Psi_{0}\vert T^G\vert\Psi_{0}\rangle+E_{eHxc}[\rho]+\sum_{i;\Gamma}E_{i\Gamma}m_{i\Gamma}-E_{dc}
\end{split}
\end{equation}
The total energy is a functional of uncorrelated wave functions
$\vert\Psi_{0}\rangle$ and the configuration weight $m_{i\Gamma}$,
and have just the same form as Eq.~(\ref{eq:totalenergy_G}). Both
should be variationally optimized as already shown in the last
section.

For a summary of this section, we have performed a explicit derivation
just in the spirit of the Kohn-Sham ansatz to incorporating the
Gutzwiller approach into density functional theory. With some
reasonable approximations, we proof that this scheme gives the same
result as what we developed in the last section within a simple
physical interpretation. This provides a reliable foundation of this
new LDA+G method.

\subsection{Interaction and Double Counting Terms}

Similar to all kinds of LDA+U or LDA+DMFT calculations, the
interaction and the double-counting terms still remain to be defined
explicitly in the present LDA+G method, it is therefore also a kind of
semi-empirical {\it ab initio} method in this sense. Nevertheless, we
can in general follow the same definition used in LDA+U or LDA+DMFT
method.  From the physical point of view, we only consider the strong
on-site interactions of localized orbitals. The interaction strength
can be explicit expressed with Slater integrals (or called
Slater-Condon parameters) ($F^0,F^2,\cdots$) in the atomic
limit. However, in practice, it is more convenient to use ``Kanamori
parameters'', $U$,$U'$,$J$, and $J'$, which are combinations of Slater
integrals. The general form of the on-site interactions can be written
as \cite{Hotta}
\begin{equation}
  \label{eq:Hi_explicit}
\begin{split}
  H_i=&U\sum_{\alpha}n_{i\alpha\uparrow}n_{i\alpha\downarrow}+\frac{U'}{2}\sum_{\alpha\neq\alpha',\chi\chi'}n_{i\alpha\chi}n_{i\alpha'\chi'} \\
  &-\frac{J}{2}\sum_{\alpha\neq\alpha',\chi}n_{i\alpha\chi}n_{i\alpha'\chi}\\
  &-\frac{J}{2}\sum_{\alpha\neq\alpha',\chi}c^{\dagger}_{i\alpha\chi}c_{i\alpha\bar{\chi}}c^{\dagger}_{i\alpha'\bar{\chi}}c_{i\alpha'\chi}\\
  &-\frac{J'}{2}\sum_{\alpha\neq\alpha'}c^{\dagger}_{i\alpha\uparrow}c^{\dagger}_{i\alpha\downarrow}c_{i\alpha'\uparrow}c_{i\alpha'\downarrow}\\
\end{split}
\end{equation}
where $\alpha$ denotes localized orbital and $\chi$ denotes spin. The
first two terms are intra-orbital Coulomb interaction, and
inter-orbital Coulomb interaction, respectively.  The Hund's rule
exchange coupling are divided into three parts: one is the
longitudinal part (the third term) which only involves density-density
coupling; the other two terms (the 4-th and 5-th terms) describe the
spin flip and pair hopping processes respectively.  In the atomic
case, the relation $U=U'+J+J'$ holds to retain the rotational
invariance in orbital space. And for typical $d$-orbital systems,
where spin-orbital coupling is not so strong, the relation $J=J'$ also
holds, we therefore have $U'=U-2J$ in general.

In this article, we restricted ourselves to the pure density-density
interactions for simplicity, and the interactions to be considered are
\begin{equation}
  \label{eq:Hi_dd}
\begin{split}
    H_i=&U\sum_{\alpha}n_{i\alpha\uparrow}n_{i\alpha\downarrow}+\frac{U'}{2}\sum_{\alpha\neq\alpha',\chi\chi'}n_{i\alpha\chi}n_{i\alpha'\chi'} \\
  &-\frac{J}{2}\sum_{\alpha\neq\alpha',\chi}n_{i\alpha\chi}n_{i\alpha'\chi}
\end{split}
\end{equation}
This on-site interaction Hamiltonian is already diagonal in the atomic
configuration $|\Gamma\rangle$ space, and the corresponding
configuration energy $E_{\Gamma}$ is a linear combination of $U$, $U'$ and
$J$~\cite{multiband-G}.

With the on-site interactions determined, we come to the question how
much of them are taken into account in LDA, that is, how to write down
the double counting terms. It is known that these interactions goes
into LDA in an average way without orbital dependence.  As already
discussed in LDA+U or LDA+DMFT methods, we just follow the common
choice of the double counting terms as~\cite{LDAU}:
\begin{equation}
\begin{split}E_{dc}[n_{i}]= & \sum_{i}\frac{\overline{U}}{2}n_{i}(n_{i}-1)\\
 & -\sum_{i}\frac{\overline{J}}{2}(n_{i\uparrow}(n_{i\uparrow}-1)+n_{i\downarrow}(n_{i\downarrow}-1))
\end{split}
\label{eq:E_dc}\end{equation}
where $n_{i}$ is the total electron number of localized orbitals on
the same site $i$,
$n_i=n_{i\uparrow}+n_{i\downarrow}=\sum_{\alpha\chi}n_{i\alpha\chi}$,
and $\overline{U}$, $\overline{J}$ are spherically averaged
interactions, which can be given as~\cite{DMFT},
\begin{equation}
\label{eq:Uavg}
  \overline{U}=\frac{1}{(2l+1)}(U+2lU')
\end{equation}

\begin{equation}
\label{eq:Javg}
  \overline{J}=\overline{U}-U'+J
\end{equation}
where $l$ is the angular momentum number of the corresponding
localized shell.

The Coulomb and exchange interactions, $U$ and $J$, can in principle
be obtained using corresponding Slater integrals. However, in real
materials, the bare electron-electron interactions must be screened,
and the Slater integrals have to be renormalized. Therefore, it is a
hard task to determine the effective $U$ and $J$ exactly. In practice,
usually two possible ways are followed: (1) determining the parameters
from available experimental information empirically; (2) calculating
the parameters from constrained LDA method~\cite{LDAU} and the linear
response approach~\cite{linearLDAU}. Depending on different method,
different values might be obtained, however the important strategy is
that for single fixed parameter, the method should be able to explain
all possible properties spontaneously and systematically, rather than
using different parameters for different properties. It is only in
this way, the obtained results can be justified. We should also notice
that the interaction parameters also depend on the choice of local
orbitals, because of the different screening processes involved. For
example, both atomic orbitals and wannier functions can be used to
define the local orbitals, however generally the effective interaction
strength for atomic orbitals should be larger than that for wannier
orbitals because the former is more localized. To construct the
wannier orbitals, either the projected wannier
method~\cite{projected_wannier} or the maximally localized wannier
function~\cite{mlwf} can be used.

Finally, a very similar approach has been recently proposed
independently by K. M. Ho. et.al. to combine the Gutzwiller approach
with the DFT \cite{gw_DFT}. The spirit of our method and their
proposal are almost the same, however, they differs in the definition
of the interaction and double counting terms. It is still remained to
be justified which way should be the best in future.

\section{Results and Discussions for Realistic Systems}

The above proposed LDA+G method was implemented in our BSTATE (Beijing
Simulational Tool for Atom Technology) code~\cite{STATE-review}, which
uses plane-wave ultra-soft pseudo-potential method, with projected
wannier function for the definition of local orbitals.  We have
applied the method to study several typical systems, where the strong
e-e interactions play important roles. They are non-magnetic metal
SrVO$_3$, magnetic metal Fe and Ni, AF insulator NiO, and
unconventional superconductor Na$_{1-x}$CoO$_2$. Some of the results
have been published~\cite{LDA+Gutzwiller, NaCoO} with emphasis on
particular issues in each example. On the other hand, the main purpose
of this full paper is to present the whole formalism of LDA+G and to
demonstrate its advantage, namely what knowledge can be gained beyond
LDA or LDA+U. Therefore, to keep the completeness of our present
paper, here we would like to concentrate on the physical consequence
of our method by grouping all results together. We will discuss all
those results in a totally different manner, such that we can
understand the LDA+G method better.

\begin{figure}[tbp]
\includegraphics[clip,width=7cm]{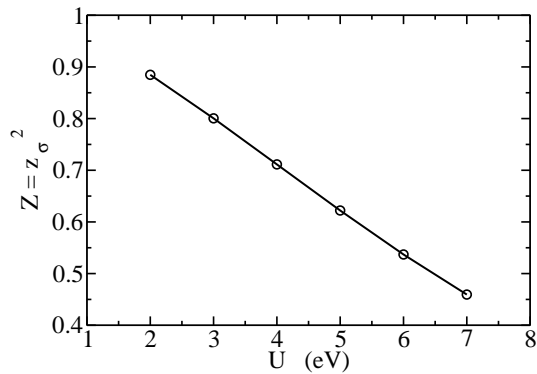}
\caption{The calculated $Z$-factor as function of $U$ for SrVO$_3$
  using LDA+G method.}
\label{fig:SrVO3-Z}
\end{figure}

\begin{figure}[tbp]
\includegraphics[clip,width=7cm]{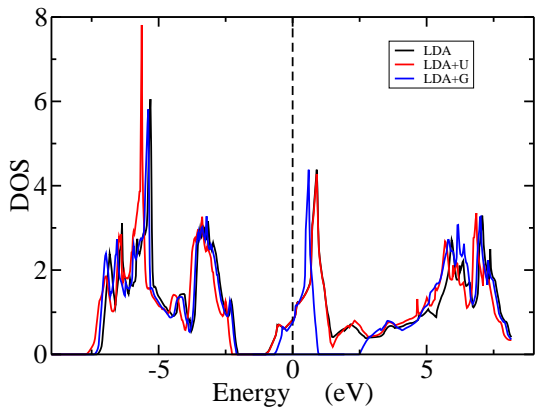}
\caption{The calculated density of states for SrVO$_3$ using LDA,
  LDA+U and LDA+G methods. We used the $U$=5.0eV and $J$=1.0eV, and
  the electron DOS (rather than quasi-particle DOS) is shown, in the
  calculations of LDA+G.}
\label{fig:SrVO3-dos}
\end{figure}

\noindent {\it 1. Band-narrowing and mass renormalization}

An essential quantity included in the LDA+G method is the kinetic
renormalization factor $Z=z^2_{\sigma}$ due to the dynamic
correlation. The $z_{\sigma}$ factors are orbital-dependent, and can
be self-consistently obtained from the energy minimization. Under the
Gutzwiller approximation, the $Z$ factor can be also understood as the
quasi-particle weight. It has
been widely recognized that LDA type calculations overestimate the
band-width of correlated-electron systems. The error-bar could be as
large as an order (such as in Heavy Fermion system)
depending on the strength of correlation. We will show here, that this
band-narrowing (or mass renormalization) physics can be correctly
obtained from the LDA+G method. For example, SrVO$_3$ is a
intermediately correlated metal with $3d$-$t_{2g}^1$ configuration. It
has simple cubic perovskite crystal structure, and magnetic
instabilities are not involved for the ground state
property~\cite{SVO-theory}. Although the LDA calculation can correctly
predict the non-magnetic metallic nature of the ground state, the
calculated band width is about 40\% wider than photoemission
observation~\cite{SVO-exp}, and the estimated effective mass is about
2-3 times lower than experimental results from specific heat and
susceptibility. On the other hand, all these features can be improved
by LDA+G calculations, and correct band-narrowing and mass
renormalization can be obtained as shown in Fig.~\ref{fig:SrVO3-Z} and
Fig.~\ref{fig:SrVO3-dos} by using reasonable $U$($\sim$ 5.0eV). To
gain further understanding, here we would also like to compare the
results to that obtained by LDA+U method. In the LDA+U, only the
interaction energy part is corrected over LDA, and the kinetic part is
not renormalized, as already discussed in the previous section.
 Therefore, LDA+U can not explain the observed large
renormalization as shown in Fig.~\ref{fig:SrVO3-dos}, where the
density of states (DOS) obtained by LDA+U almost coincide with that by
LDA.

 Finally, we will also show below that the band-narrowing
and mass renormalization is such an important quantity and common
advantage of LDA+G that it is encountered for all the examples we
studied.

\begin{figure}[tbp]
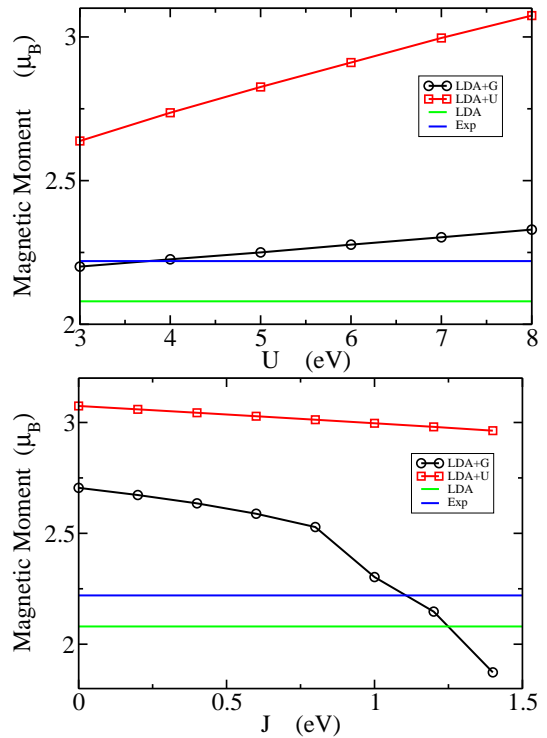

\includegraphics[clip,width=7cm]{Fe_M_U_vs}
\includegraphics[clip,width=7cm]{Fe_M_J_vs}
\caption{The calculated magnetic moment (per Fe) of bcc FM Fe as
  function of $U$ and $J$, by using different methods.}
\label{fig:Fe-xc}
\end{figure}

\begin{figure}[tbp]
\includegraphics[clip,width=7cm]{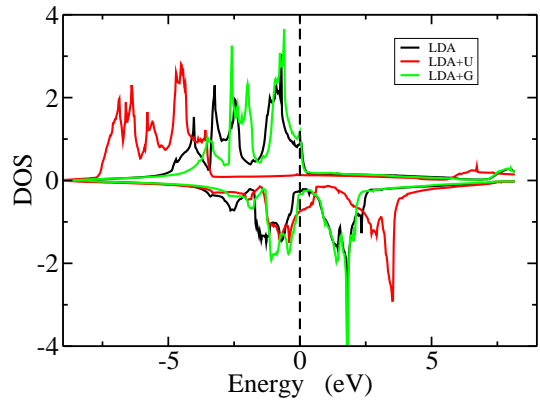}
\caption{The calculated DOS of bcc FM Fe using different methods. The
  parameters $U$=7.0eV and $J$=1.0eV are used in these calculations.}
\label{fig:Fe-dos}
\end{figure}

\noindent{\it 2. Improved spin polarization}

Except the band renormalization, we will show here that the spin
polarization of magnetic systems can be also improved to certain
extend. To understand the physics better, we would like to divide the
mechanisms of spin polarization into two parts: (1) The inter-site
exchange (or the spacial long-range exchange); (2) The intra-site
exchange (mostly the inter-orbital Hund's coupling).  Such a
separation is not rigorous, but just for the physical
understanding. It is important to note that, in our formalism, the
spacial inter-site part still remain to be treated by LDA level, and
only the intra-site part is improved explicitly. Of course, through
the charge-density self-consistency, the inter-site part may be also
tuned slightly, but it is not a main effect. It is therefore
understood that the issues related to the inter-site exchange, such as
the spin spacial fluctuation or the geometrically frustrated spin
systems, can not be improved through the LDA+G treatment.

Even for the intra-site interaction, it is treated both in LDA+U and
in LDA+G, what will be the difference? In the LDA+U, it is treated
from the static mean field level, which always tends to give larger
spin polarization than that in LDA for positive effective
$U_{eff}$(=$U$-$J$). On the other hand in the LDA+G method, the
dynamic effects are included and the intra-site (inter-orbital) charge
and spin fluctuations are all included in a better way. It is in this
sense that the results by LDA+G should be more reasonable.  To
demonstrate the effect of LDA+G on spin polarization, here we would
like to show three examples:

(1) For ferromagnetic (FM) bcc Fe, as shown in Fig.~\ref{fig:Fe-xc}
and \ref{fig:Fe-dos}, the calculated magnetic moment in LDA+U is
always larger than that in LDA, and is significantly overestimated
compared to experiments. On the other hand, in LDA+G, this
overestimation by LDA+U is suppressed. (2) For the bulk fcc Ni, the
calculated moment by LDA+G is even smaller than that in LDA, in better
agreement with experiments (see Table I).  (3) For Na$_{1-x}$CoO$_2$,
the LDA level calculations predict that the system is magnetic for all
the doping range, the LDA+U calculations even enhance the tendency to
be magnetic, in contrast to experimental observation. However, using
LDA+G, we show in Ref.~\cite{NaCoO} that the ground state is actually
non-magnetic for the intermediate doping range (around 0.3$<x<$0.5),
in nice agreement with experiments.

\noindent {\it 3. Total energy and equilibrium properties}

A big advantage of LDA type calculations based on DFT is its ability
to get the ground state total energy accurately. Here we will show
that, by explicitly treating the interaction term through our LDA+G
formalism, the calculated total energy and equilibrium properties of
correlated electron systems can be also improved significantly.  Bulk
Fe and Ni are typical magnetic metals with intermediate correlations,
where LDA produce big error bar for the ground state properties in
comparison to experiments.  For Fe, the LDA even fails to predict the
correct bcc FM ground state (although GGA correctly do so, the reason
is not clearly understood). The results, summarized in Table I shows
that most of the discrepancies are systematically improved, compared
with experiments, suggesting the advantages of present scheme. First
of all, the bcc FM ground state is now correctly predicted by LDA+G
(see Ref.~\cite{LDA+Gutzwiller} for original figure), we therefore
understand that the failure of LDA to predict the correct ground state
is due to its underestimation of strong on-site correlation. Second,
the calculated equilibrium volume, bulk modulus, magnetic moment,
specific heat coefficient, and band width are all improved in a
systematic way by a simple fixed interaction strength ($U$=7.0eV and
$J$=1.0eV). This is in sharp contrast to that obtained in LDA+U, for
instance, the LDA+U may also get the correct equilibrium volume by
certain $U$ value, but the obtained magnetic moment will be
unreasonably larger than experimental results if the same $U$ is used.

\begin{table}[tbp]
  \caption{The calculated property parameters for bcc FM Fe and fcc FM Ni in
    comparison with experimental results. They are equilibrium lattice constant $
    a_0$, bulk modulus $B$, spin magnetic moment $M$, specific heat coefficient $
    \protect\gamma$, and the occupied energy band width $W$. The experimental
    data are from Ref.~\protect\cite{Fe-Ni-Exp}. 
    (This table is a reproduction of our results that published in Ref~\cite{LDA+Gutzwiller}.)}
\begin{center}
\scalebox{0.8}{\begin{tabular}{c|l|c|c|c|c|c}
\hline
     &      &a$_0$(bohrs)  &$B$(GPa)   &$M$($\mu_B$) &$\gamma$($\frac{mJ}{k^2mol}$) &$W$(eV) \\ \hline
     &LDA   &5.21          &227      &2.08       &2.25           &3.6    \\
Fe   &LDA+G &5.39          &160      &2.30       &3.52           &3.2    \\
     &Exp.  &5.42          &168      &2.22       &3.1,3.69       &3.3     \\
\hline
     &LDA   &6.49          &250      &0.59       &4.53           &4.5     \\
Ni   &LDA+G &6.61          &188      &0.50       &6.9            &3.2     \\
     &Exp.  &6.65          &186      &0.42,0.61  &7.02           &3.2    \\  \hline
\end{tabular}
}
\end{center}
\end{table}

\noindent {\it 4. Large gap AF ordered Insulator}

Now we come to discussions for the large gap AF ordered insulator with
integer occupation, where LDA+U works well. We will show that the
LDA+G actually gives similar results in this limit. The reason is very
straight forward as has been pointed out in the formalism.  In the
present LDA+G scheme, both the on-site level and the kinetic energy
should be renormalized due to the presence of interaction term.
However, in the case that long-range ordering is established with
integer occupation, if the energy gap is big, each orbital should be
close to either fully occupied or totally empty, because the charge
fluctuation between the states should be small. In this limit, the
kinetic renormalization $Z$ factor will be very close to unity, and
$\Psi_G$ returns back to $\Psi_0$. Therefore, the kinetic
renormalization is very small, and only the renormalization to the
on-site level take effect, this is exactly just the limit that
obtained in LDA+U. As we have shown in the calculations for
NiO~\cite{LDA+Gutzwiller}, the obtained electronic structure is very
similar to that of LDA+U. However please note, even for the AF
long-range ordered insulators, if the band gap is small and the spin
moment is far away from integer, the dynamic processes crossing the
band gap may also take effect, in this case, the $Z$ factor will be no
longer unity, and of course, the results by LDA+G will be different
with that of LDA+U, and the one by LDA+G should be more close to
reality. We expect that this situation may happen in the
LaTiO$_3$~\cite{LTO}, where the gap is about 0.2eV and the calculated
moment from LDA+U is much larger than that observed experimentally.

\noindent {\it 5. Effect of charge density self-consistency}

\begin{figure}[tbp]
\includegraphics[clip,width=7cm]{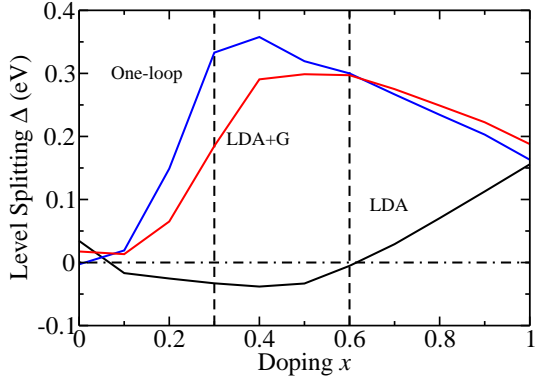}
\caption{The calculated level splitting between the $a_{1g}$ and the
  $e_{g}^\prime$ states for Na$_{1-x}$CoO$_2$. The original figure is
  obtained from Ref.~\cite{NaCoO}. The one-loop result corresponds to
  the calculation without charge density self-consistency.}
\label{fig:NaCoO-sp}
\end{figure}

Here we will show that the charge density self-consistency is really
important for the calculations on realistic systems, and we take
Na$_{1-x}$CoO$_2$, a typical multi-orbital system, as an example. As
we have pointed out in our recent publication~\cite{NaCoO}, all the
issues discussed above, such as the band-narrowing, spin
polarizations, orbital fluctuations, are encountered in
Na$_{1-x}$CoO$_2$, and systematic improvement are obtained through
LDA+G treatment. However, we want to take Na$_{1-x}$CoO$_2$ as an
example to demonstrate the importance of charge self-consistency,
because several post-LDA techniques (without charge self-consistency)
have been applied to this compound and conflicting results are
obtained~\cite{GW_NaCoO,DMFT_NaCoO}. The issue is related to the
relative splitting of energy level between $e_g^\prime$ and $a_{1g}$
states, and the appearance of $e_g^\prime$ hole pockets at the Fermi
surface (for $x$=0.3). If the splitting is large, the $e_g^\prime$
orbital will be totally occupied, and there will be no $e_g^\prime$
hole pockets at the Fermi surface.  Starting from different
Hamiltonians fitted to LDA band structure, advanced techniques such as
Gutzwiller or DMFT have been applied in post-LDA scheme, however, one
of the results suggests the absence of $e_g^\prime$ hole
pockets~\cite{GW_NaCoO}, and the other suggests the
appearance~\cite{DMFT_NaCoO}. It is now understood~\cite{DMFT_NaCoO_2}
that the main reason is due to the difference in the fitted
tight-binding Hamiltonian, namely the crystal field splitting (or the
on-site energy) in the two studies are different. In our LDA+G method,
full charge self-consistency is achieved and no tight-binding fitting
is required. Only after such kind treatment, the discrepancy can be
nicely resolved~\cite{NaCoO}. On the other hand, because of the
feedback effect in the charge self-consistency, the on-site level
renormalization which is overestimated by post-LDA techniques is now
suppressed as show Fig.~\ref{fig:NaCoO-sp}.

In summary, we have show in this full paper the detailed formalism of
LDA+Gutzwiller method, and its firm derivation from the GDFT. By
comparing the results to that obtained by DMFT, we have shown that the
energy resolution of Gutzwiller approach is pretty good for the ground
state determination. It is computationally cheaper, and yet with
dynamic fluctuations included. The calculated results for several
typical systems demonstrate that it can be widely applied to many of
the correlated electron systems with quality beyond LDA+U.

\acknowledgments

We acknowledge valuable discussions with Prof. Lu Yu, Y. G. Yao, and
the support from NSF of China and the 973 program of China.

\ \\
\  \\

\appendix*

\makeatletter
\renewcommand\theequation{A.\@arabic\c@equation }
\makeatother \setcounter{equation}{0}

\begin{center}
    {\bf APPENDIX}
\end{center}

In this appendix, we will proof several equations discussed in the
text part. We will pay special attention to when and how GA is
applies. Even a variational ansatz is given for a lattice model,
evaluation of expect value is not straightforward. GA is a
systemically approximation to make the evaluation accessible. The
spirit of GA actually is to neglect Wick contractions between
operators with different site/orbital indices, thus in the following
we often need to Fourier transform the expression into real space and
then apply GA.

\subsection{Proof of Eqn. \protect\ref{eq:mgutzH}}

First we note that by choose $\lambda _{i\Gamma }=\sqrt{\frac{m_{i\Gamma }}{%
m_{i\Gamma }^{0}}}$, $|\Psi _{G}\rangle $ is normalized under GA. $\langle
\Psi _{G}|\Psi _{G}\rangle =\prod\limits_{i}\langle \Psi _{0}|\hat{P}%
_{i}^{2}|\Psi _{0}\rangle =\prod\limits_{i}\sum_{\Gamma }\frac{m_{i\Gamma }}{%
m_{i\Gamma }^{0}}\langle \Psi _{0}|\hat{m}_{i;\Gamma }|\Psi _{0}\rangle
=\prod\limits_{i}\left( \sum_{\Gamma }m_{i\Gamma }\right) =1$. In the first
equality we separate the average of a projection operator string into the
product of single site averages.

Then the expectation value of kinetic energy is
\begin{align*}
  & \langle \Psi _{G}|H_{0}|\Psi _{G}\rangle \\
  & =\sum_{i,j}t_{ij}^{\sigma \sigma ^{\prime }}\langle \Psi
  _{0}|\hat{P} _{i}C_{i\sigma }^{\dagger
  }\hat{P}_{i}\hat{P}_{j}C_{j\sigma ^{\prime }}\hat{%
    P }_{j}|\Psi _{0}\rangle \times \langle \Psi
  _{0}|\prod\limits_{i^{\prime
    }\neq i,j}\hat{P}_{i^{\prime }}^{2}|\Psi _{0}\rangle \\
  & =\sum_{i,j}\sum_{\substack{ \Gamma _{i},\Gamma _{i}^{^{\prime }}
      \\ %
      \Gamma _{j},\Gamma _{j}^{^{\prime }}}}t_{ij}^{\sigma \sigma
    ^{\prime }}\sqrt{ \frac{m_{\Gamma _{i}}m_{\Gamma _{i}^{^{\prime
          }}}m_{\Gamma _{j}}m_{\Gamma _{j}^{^{\prime }}}}{m_{\Gamma
        _{i}}^{0}m_{\Gamma _{i}^{^{\prime
          }}}^{0}m_{\Gamma _{j}}^{0}m_{\Gamma _{j}^{^{\prime }}}^{0}}} \\
  & \qquad \times \langle \Psi _{0}|\hat{m}_{i;\Gamma _{i}^{^{\prime
      }}}C_{i\sigma }^{\dagger }\hat{m}_{i;\Gamma
    _{i}}\hat{m}_{j;\Gamma _{j}}C_{j\sigma ^{\prime
    }}\hat{m}_{j;\Gamma _{j}^{^{\prime }}}|\Psi _{0}\rangle
\end{align*}
where in the first equality we adopt GA to neglect all Wick
contractions from site $i^{\prime }\neq i,j$ and site $i,j$. To
evaluate the expectation value we define
\begin{eqnarray*}
\hat{m}_{i;\Gamma _{i}^{^{\prime }}} &=&\hat{l}_{i;\Gamma _{i}}\hat{n}
_{i\sigma } \\
\hat{m}_{i;\Gamma _{i}} &=&\hat{l}_{i;\Gamma _{i}}\left( 1-\hat{n}_{i\sigma
}\right) \\
\hat{m}_{j;\Gamma _{j}^{^{\prime }}} &=&\hat{l}_{_{j;\Gamma _{j}}}\hat{n}
_{j\sigma ^{\prime }} \\
\hat{m}_{j;\Gamma _{j}} &=&\hat{l}_{j;\Gamma _{j}}\left( 1-\hat{n}_{j\sigma
\prime }\right)
\end{eqnarray*}
where $\hat{l}_{i;\Gamma _{i}}$ ( $\hat{l}_{j;\Gamma _{j}}$) are
projection operators for orbital other than $\sigma $($\sigma ^{\prime
}$). Then we have
\begin{align}
  & \qquad \langle \Psi
  _{0}|\mathcal{\hat{P}}H_{0}\mathcal{\hat{P}}|\Psi
  _{0}\rangle  \notag \\
  =& \sum_{i,j}\sum_{_{\substack{ \Gamma _{i},\Gamma _{i}^{^{\prime }}
        \\ %
        \Gamma _{j},\Gamma _{j}^{^{\prime }}}}ij}t_{ij}^{\sigma \sigma
    ^{\prime }} \sqrt{m_{\Gamma _{i}}m_{\Gamma _{i}^{^{\prime
        }}}m_{\Gamma _{j}}m_{\Gamma _{j}^{^{\prime }}}}\frac{\langle
    \Psi _{0}|\hat{l}_{i;\Gamma _{i}}\hat{l} _{i;\Gamma
      _{i}}\hat{l}_{j;\Gamma _{j}}\hat{l}_{j;\Gamma _{j}}|\Psi
    _{0}\rangle }{\sqrt{m_{\Gamma _{i}}^{0}m_{\Gamma _{i}^{^{\prime
          }}}^{0}m_{\Gamma _{j}}^{0}m_{\Gamma _{j}^{^{\prime }}}^{0}}}  \notag \\
  & \times \langle \Psi _{0}|\hat{n}_{i\sigma }C_{i\sigma }^{\dagger
  }\left( 1- \hat{n}_{i\sigma }\right) \left( 1-\hat{n}_{j\sigma
      ^{\prime }}\right) C_{j\sigma ^{\prime }}\hat{n}_{j\sigma
    ^{\prime }}|\Psi _{0}\rangle D_{\Gamma _{i}^{^{\prime }}\Gamma
    _{i}}^{\sigma }D_{\Gamma _{j}^{^{\prime
      }}\Gamma _{j}}^{\sigma ^{\prime }\ast }  \notag \\
  =& \sum_{i,j}\sum_{\substack{ \Gamma _{i},\Gamma _{i}^{^{\prime
        }}\Gamma _{j},\Gamma _{j}^{^{\prime }}}}t_{ij}^{\sigma \sigma
    ^{\prime }}\frac{\sqrt{ m_{\Gamma _{i}}m_{\Gamma _{i}^{^{\prime
          }}}}D_{\Gamma _{i}^{^{\prime }}\Gamma _{i}}^{\sigma
    }}{\sqrt{n_{i\sigma }^{0}\left( 1-n_{i,\sigma }^{0}\right)
    }}\frac{\sqrt{m_{\Gamma _{j}}m_{\Gamma _{j}^{^{\prime }}}}
    D_{\Gamma _{j}^{^{\prime }}\Gamma _{j}}^{\sigma ^{\prime }\ast
    }}{\sqrt{
      n_{j\sigma }^{0}\left( 1-n_{j,\sigma }^{0}\right) }}  \notag \\
  & \times \langle \Psi _{0}|C_{i\sigma }^{\dagger }C_{j\sigma
    ^{\prime
    }}|\Psi _{0}\rangle  \notag \\
  =& \sum_{i,j,\sigma }z_{i\sigma }z_{j\sigma ^{\prime
    }}t_{ij}^{\sigma \sigma ^{\prime }}\langle \Psi _{0}|C_{i\sigma
  }^{\dagger }C_{j\sigma ^{\prime }}|\Psi _{0}\rangle \label{kin}
\end{align}

The expectation value of interaction part is
\begin{align}
  & \langle \Psi _{G}|H_{int}|\Psi _{G}\rangle  \notag \\
  & =\sum_{i}\sum_{\Gamma }E_{\Gamma }\frac{m_{i;\Gamma }}{m_{i;\Gamma
    }^{0}}
  \langle \Psi _{0}|\hat{m}_{i;\Gamma }|\Psi _{0}\rangle  \notag \\
  & =\sum_{i}\sum_{\Gamma }E_{\Gamma }m_{i;\Gamma } \label{int}
\end{align}

Put equations \ref{kin} and \ref{int} together, we will have the
equation~(\ref{eq:mgutzH}) show in the text part.

\subsection{\protect\bigskip Proof of $\protect\gamma _{k\protect\sigma %
}=z_{ \protect\sigma }$}

\bigskip In this section, we proof $\gamma _{k\sigma }=z_{\sigma }$ under
GA, here we focus on quasi-particle sector without loss of generality. To
compute $<\Phi _{k\sigma }^{p}|C_{k\sigma }^{\dagger }|G>$, we first Fourier
transform the operators into real space then apply GA.

\begin{align*}
& \langle \Phi _{k\sigma }^{p}|C_{k\sigma }^{\dagger }|\Psi _{G}\rangle \\
& =\frac{1}{N}\sum_{I,J}e^{ik\left( I-J\right) }\langle \Psi _{0}|C_{J\sigma
}\mathcal{\hat{P}}C_{I\sigma }^{\dagger }\mathcal{\hat{P}}|\Psi _{0}\rangle
\\
& =\frac{1}{N}\sum_{I}\langle \Psi _{0}|C_{I\sigma }\mathcal{\hat{P}}
C_{I\sigma }^{\dagger }\mathcal{\hat{P}}|\Psi _{0}\rangle \\&+\frac{1}{N}\sum
_{\substack{ I,J  \\ I\neq J}}e^{ik\left( I-J\right) }\langle \Psi
_{0}|C_{J\sigma }\mathcal{\hat{P}}C_{I\sigma }^{\dagger }\mathcal{\hat{P}}
|\Psi _{0}\rangle \\
& =\frac{z_{\sigma }}{N}\sum_{I}\langle \Psi _{0}|C_{I\sigma }C_{I\sigma
}^{\dagger }|\Psi _{0}\rangle \\ & +\frac{z_{\sigma }}{N}\sum_{\substack{ I,J  \\ %
I\neq J}}e^{ik\left( I-J\right) }\langle \Psi _{0}|C_{J\sigma }C_{I\sigma
}^{\dagger }|\Psi _{0}\rangle \\
& =z_{\sigma }\langle \Psi _{0}|C_{k\sigma }C_{k\sigma }^{\dagger }|\Psi
_{0}\rangle \\
& =z_{\sigma }\theta (\varepsilon _{k\sigma }-\mu _{F})
\end{align*}

\bigskip Similarly, we could proof that under GA $\langle \Phi
_{k\sigma }^{h}|C_{k\sigma }|\Psi _{G}\rangle =z_{\sigma }\langle \Psi
_{0}|C_{k\sigma }^{\dagger }C_{k\sigma }|\Psi _{0}\rangle =z_{\sigma
}\theta (\mu _{F}-\varepsilon _{k\sigma })$. Thus we have $\gamma
_{k\sigma }=z_{\sigma }$ under GA. Also note that under GA the $k$
dependence of the $Z$ is missing, only the fact that above or below
the Fermi surface matters.

\subsection{\protect\bigskip Evaluation of $E_{k\protect\sigma }^{p}$}

First we show that the quasi-particle state is normalized under GA:

\begin{eqnarray*}
&&\langle \Phi _{k\sigma }^{p}|\Phi _{k\sigma }^{p}\rangle \\
&=&\frac{1}{N}\sum_{I}\sum_{\Gamma }\frac{m_{\Gamma }}{m_{\Gamma }^{0}}
\langle \Psi _{0}|C_{I\sigma }\hat{m}_{I;\Gamma }C_{I\sigma }^{\dagger
}|\Psi _{0}\rangle+ \\
&&\frac{1}{N}\sum_{\substack{ I,J  \\ I\neq J}}e^{ik\left( I-J\right)
}\sum_{\Gamma _{I,}\Gamma _{J}}\frac{m_{\Gamma _{I}}}{m_{\Gamma _{I}}^{0}}
\frac{m_{\Gamma _{J}}}{m_{\Gamma _{J}}^{0}}\langle \Psi _{0}|C_{J\sigma }
\hat{m}_{J;\Gamma _{J}}\hat{m}_{I;\Gamma _{I}}C_{I\sigma }^{\dagger }|\Psi
_{0}\rangle \\
&=&\frac{1}{N}\sum_{I}\langle \Psi _{0}|C_{I\sigma }C_{I\sigma }^{\dagger
}|\Psi _{0}\rangle +\frac{1}{N}\sum_{\substack{ I,J  \\ I\neq J}}e^{ik\left(
I-J\right) }\langle \Psi _{0}|C_{J\sigma }C_{I\sigma }^{\dagger }|\Psi
_{0}\rangle \\
&=&\langle \Psi _{0}|C_{k\sigma }C_{k\sigma }^{\dagger }|\Psi _{0}\rangle \\
&=&1\text{ for }\varepsilon _{k\sigma }>\mu _{F}
\end{eqnarray*}

Then the kinetic energy for spin $\sigma $ species reads,

\begin{align*}
& \langle \Phi _{k\sigma }^{p}|\sum_{i,j}t_{ij}C_{i\sigma }^{\dagger
}C_{j\sigma }|\Phi _{k\sigma }^{p}\rangle \\
& =\frac{1}{N}\sum_{I,J}e^{ik\left( I-J\right) }\sum_{i,j}t_{ij}\langle \Psi
_{0}|C_{J\sigma }\mathcal{\hat{P}}C_{i\sigma }^{\dagger }C_{j\sigma }
\mathcal{\hat{P}}C_{I\sigma }^{\dagger }|\Psi _{0}\rangle \\
& =\sum_{i,j}t_{ij}\langle \Psi _{0}|C_{I\sigma }\mathcal{\hat{P}}C_{i\sigma
}^{\dagger }C_{j\sigma }\mathcal{\hat{P}}C_{I\sigma }^{\dagger }|\Psi
_{0}\rangle \\
& +\sum_{J\neq I}e^{ik\left( I-J\right) }\sum_{i,j}t_{ij}\langle \Psi
_{0}|C_{J\sigma }\mathcal{\hat{P}}C_{i\sigma }^{\dagger }C_{j\sigma }
\mathcal{\hat{P}}C_{I\sigma }^{\dagger }|\Psi _{0}\rangle
\end{align*}

For the first term since $i\neq j,$ there is a constrain that $I\neq i$ and $%
I\neq j$, other wise the expression vanishes. Then it equals to

\begin{eqnarray*}
&&\sum_{i,j}t_{ij}\langle \Psi _{0}|C_{I\sigma }\hat{P}_{I}\hat{P}_{i}\hat{P}
_{j}C_{i\sigma }^{\dagger }C_{j\sigma }\hat{P}_{i}\hat{P}_{j}\hat{P}
_{I}C_{I\sigma }^{\dagger }|\Psi _{0}\rangle \\
&=&(\sum_{\substack{ \Gamma _{I}\supset \sigma }}\frac{m_{\Gamma _{I}}}{
n_{\sigma }^{0}})z_{\sigma }^{2}\sum_{i,j}t_{ij}\langle \Psi _{0}|C_{I\sigma
}C_{i\sigma }^{\dagger }C_{j\sigma }C_{I\sigma }^{\dagger }|\Psi _{0}\rangle
\\
&=&z_{\sigma }^{2}\sum_{i,j}t_{ij}\langle \Psi _{0}|C_{I\sigma }C_{i\sigma
}^{\dagger }C_{j\sigma }C_{I\sigma }^{\dagger }|\Psi _{0}\rangle
\end{eqnarray*}
For the second term, we have $4$ cases: $I=j$ but $J\neq i$; $I\neq j$ but $%
J=i$; $I=j$, $J=i$ and $I\neq j,J\neq i$. Following previous technique, one
could find out each of the cases the projection operator $\mathcal{\hat{P}}\
$gives a $z_{\sigma }^{2}$ factor. Thus,

\begin{eqnarray*}
&<&\Phi _{k\sigma }^{p}|\sum_{i,j}t_{ij}C_{i\sigma }^{\dagger }C_{j\sigma
}|\Phi _{k\sigma }^{p}> \\
&=&z_{\sigma }^{2}\sum_{i,j}t_{ij}\langle \Psi _{0}|C_{k\sigma }C_{i\sigma
}^{\dagger }C_{j\sigma }C_{k\sigma }^{\dagger }|\Psi _{0}\rangle \\
&=&z_{\sigma }^{2}\sum_{k}\varepsilon _{k\sigma }\langle \Psi
_{0}|C_{k\sigma }^{\dagger }C_{k\sigma }|\Psi _{0}\rangle +z_{\sigma
}^{2}\varepsilon _{k\sigma }
\end{eqnarray*}

While kinetic energy for $\mu \neq \sigma $:

\begin{align*}
& \langle \Phi _{k\sigma }^{p}|\sum_{i,j,\mu }t_{ij}C_{i\mu }^{\dagger
}C_{j\mu }|\Phi _{k\sigma }^{p}\rangle \\
=& \sum_{J}e^{ik\left( I-J\right) }\sum_{i,j,\mu }t_{ij}\langle \Psi
_{0}|C_{J\sigma }\mathcal{\hat{P}}C_{i\mu }^{\dagger }C_{j\mu }\mathcal{\hat{
P}}C_{I\sigma }^{\dagger }|\Psi _{0}\rangle \\
=& [\langle \Psi _{0}|C_{I\sigma }C_{I\sigma }^{\dagger }|\Psi _{0}\rangle
+\sum_{J\neq I}e^{ik\left( I-J\right) }\langle \Psi _{0}|C_{J\sigma
}C_{I\sigma }^{\dagger }|\Psi _{0}\rangle ] \\
& \times z_{\mu }^{2}\sum_{i,j}t_{ij}\langle \Psi _{0}|C_{i\mu }^{\dagger
}C_{j\mu }|\Psi _{0}\rangle \\
=& \sum_{k\mu }z_{\mu }^{2}\varepsilon _{k\mu }\langle \Psi _{0}|C_{k\mu
}^{\dagger }C_{k\mu }|\Psi _{0}\rangle
\end{align*}

\bigskip Following the same routine, one could proof that

\begin{equation*}
\langle \Phi _{k\sigma }^{p}|H_{I}|\Phi _{k\sigma }^{p}\rangle
=N\sum_{\Gamma }E_{\Gamma }m_{\Gamma }+O(1)
\end{equation*}

Put the kinetic and interaction energy together, we get $\left\langle \Phi
_{k\sigma }^{p}\right\vert H\left\vert \Phi _{k\sigma }^{p}\right\rangle
=\sum_{k\mu }z_{\mu }^{2}\varepsilon _{k\mu }<0|C_{k\mu }^{\dagger }C_{k\mu
}|0>+NE_{\Gamma }m_{\Gamma }+z_{\sigma }^{2}\epsilon _{k\sigma }+O(1)$, put
the constant into chemical potential, we have
\begin{eqnarray*}
E_{k\sigma }^{p} &=&\frac{\langle \Phi _{k\sigma }^{p}|H|\Phi _{k\sigma
}^{p}\rangle }{\langle \Phi _{k\sigma }^{p}|\Phi _{k\sigma }^{p}\rangle }
-E_{G} \\
&=&z_{\sigma }^{2}\left( \epsilon _{k\sigma }-\mu _{F}\right)
\end{eqnarray*}

For quasi-hole state, one could get similar results.

\end{document}